\documentclass[twocolumn,showpacs,showkeys,preprintnumbers,amsmath,amssymb]{revtex4}
\usepackage{graphicx}
\usepackage{dcolumn}
\usepackage{bm}
\begin{document}
%%%%%%%%%%%%%%%%%%%%%%%%%%%%%%%%%%%%%%%%%%%%%%%%%%%%%%%%%%%%%%%%%%%%%%%%%%%%%%%%%%%
\hyphenation{stand-ard}
\hyphenation{spectro-meter}
\hyphenation{ex-peri-ment}
\hyphenation{re-mains}
\hyphenation{ac-count}
\hyphenation{Me-cha-nism}
\hyphenation{mo-del}
\hyphenation{RNA}
\hyphenation{pro-tons}
\hyphenation{events}
\hyphenation{se-lect}
\hyphenation{pro-cesses}
\hyphenation{in-de-pen-dent}
\hyphenation{stu-dy}
\hyphenation{bins}
\hyphenation{dis-tri-bu-tions}
\hyphenation{o-ther}
\hyphenation{runs}
\hyphenation{re-so-nan-ces}
\hyphenation{evalu-ated}
\hyphenation{nor-ma-li-za-tion}
\hyphenation{a-zi-mu-thal}
\hyphenation{at-tempt}
\hyphenation{dif-fe-rent}
\hyphenation{ex-peri-mental}
\hyphenation{ki-ne-ma-tic}
\hyphenation{re-so-lu-tion}
\hyphenation{Hall}
\hyphenation{Fi-gu-res}
\hyphenation{thre-shold}
\hyphenation{thre-sholds}
\hyphenation{ener-gy}
\hyphenation{who-le}
\hyphenation{scat-te-ring}
\hyphenation{sin-ce}
\hyphenation{theo-re-ti-cal}
\hyphenation{sca-le}
%%%%%%%%%%%%%%%%%%%%%%%%%%%%%%%%%%%%%%%%%%%%%%%%%%%%%%%%%%%%%%%%%%%%%%%%%%%%%%%%%%%%%%%%%%%%
 
 \newcounter{univ_counter}
 \setcounter{univ_counter} {0}

\addtocounter{univ_counter} {1} 
\edef\INFNFR{$^{\arabic{univ_counter}}$ } 

\addtocounter{univ_counter} {1} 
\edef\ROMA{$^{\arabic{univ_counter}}$ } 

\addtocounter{univ_counter} {1} 
\edef\ASU{$^{\arabic{univ_counter}}$ } 

\addtocounter{univ_counter} {1} 
\edef\SACLAY{$^{\arabic{univ_counter}}$ } 

\addtocounter{univ_counter} {1} 
\edef\UCLA{$^{\arabic{univ_counter}}$ } 

\addtocounter{univ_counter} {1} 
\edef\CMU{$^{\arabic{univ_counter}}$ } 

\addtocounter{univ_counter} {1} 
\edef\CUA{$^{\arabic{univ_counter}}$ } 

\addtocounter{univ_counter} {1} 
\edef\CNU{$^{\arabic{univ_counter}}$ } 

\addtocounter{univ_counter} {1} 
\edef\UCONN{$^{\arabic{univ_counter}}$ } 

\addtocounter{univ_counter} {1} 
\edef\DUKE{$^{\arabic{univ_counter}}$ } 

\addtocounter{univ_counter} {1} 
\edef\EDINBURGH{$^{\arabic{univ_counter}}$ } 

\addtocounter{univ_counter} {1} 
\edef\FIU{$^{\arabic{univ_counter}}$ } 

\addtocounter{univ_counter} {1} 
\edef\FSU{$^{\arabic{univ_counter}}$ } 

\addtocounter{univ_counter} {1} 
\edef\GWU{$^{\arabic{univ_counter}}$ } 

\addtocounter{univ_counter} {1} 
\edef\GLASGOW{$^{\arabic{univ_counter}}$ } 

\addtocounter{univ_counter} {1} 
\edef\INFNGE{$^{\arabic{univ_counter}}$ } 

\addtocounter{univ_counter} {1} 
\edef\ORSAY{$^{\arabic{univ_counter}}$ } 

\addtocounter{univ_counter} {1} 
\edef\ITEP{$^{\arabic{univ_counter}}$ }

\addtocounter{univ_counter} {1} 
\edef\JMU{$^{\arabic{univ_counter}}$ } 

\addtocounter{univ_counter} {1} 
\edef\KYUNGPOOK{$^{\arabic{univ_counter}}$ } 

\addtocounter{univ_counter} {1} 
\edef\MIT{$^{\arabic{univ_counter}}$ } 

\addtocounter{univ_counter} {1} 
\edef\UMASS{$^{\arabic{univ_counter}}$ } 

\addtocounter{univ_counter} {1} 
\edef\MSU{$^{\arabic{univ_counter}}$ } 

\addtocounter{univ_counter} {1} 
\edef\UNH{$^{\arabic{univ_counter}}$ } 

\addtocounter{univ_counter} {1} 
\edef\NSU{$^{\arabic{univ_counter}}$ } 

\addtocounter{univ_counter} {1} 
\edef\OHIOU{$^{\arabic{univ_counter}}$ } 

\addtocounter{univ_counter} {1} 
\edef\ODU{$^{\arabic{univ_counter}}$ } 

\addtocounter{univ_counter} {1} 
\edef\PITT{$^{\arabic{univ_counter}}$ } 

\addtocounter{univ_counter} {1} 
\edef\RPI{$^{\arabic{univ_counter}}$ } 

\addtocounter{univ_counter} {1} 
\edef\RICE{$^{\arabic{univ_counter}}$ } 

\addtocounter{univ_counter} {1} 
\edef\URICH{$^{\arabic{univ_counter}}$ } 

\addtocounter{univ_counter} {1} 
\edef\SCAROLINA{$^{\arabic{univ_counter}}$ } 

\addtocounter{univ_counter} {1} 
\edef\UTEP{$^{\arabic{univ_counter}}$ }

\addtocounter{univ_counter} {1} 
\edef\JLAB{$^{\arabic{univ_counter}}$ }  

\addtocounter{univ_counter} {1} 
\edef\UCONY{$^{\arabic{univ_counter}}$ } 

\addtocounter{univ_counter} {1} 
\edef\VT{$^{\arabic{univ_counter}}$ } 

\addtocounter{univ_counter} {1} 
\edef\VIRGINIA{$^{\arabic{univ_counter}}$ } 

\addtocounter{univ_counter} {1} 
\edef\WM{$^{\arabic{univ_counter}}$ } 

\addtocounter{univ_counter} {1} 
\edef\YEREVAN{$^{\arabic{univ_counter}}$ }

\title{{\large Complete Angular Distribution Measurements of\\ Two-Body  
Deuteron Photodisintegration between 0.5 and 3~GeV}}

%%%%%%%%%%%%%%%%%%%% authors %%%%%%%%% 
 \author{ 
M.~Mirazita,\INFNFR\
F.~Ronchetti,\INFNFR$^,$\ROMA\
P.~Rossi,\INFNFR\
E.~De~Sanctis,\INFNFR\ 
G.~Adams,\RPI\
P.~Ambrozewicz,\FIU\
E.~Anciant,\SACLAY\
M.~Anghinolfi,\INFNGE\
B.~Asavapibhop,\UMASS\
G.~Audit,\SACLAY
% ...
\\
(The CLAS collaboration)
} 
\address{\INFNFR Istituto Nazionale di Fisica Nucleare, Laboratori Nazionali di Frascati, PO 13, 00044 Frascati, Italy}
\address{\ROMA Universit\`{a} di ROMA III, 00146 Roma, Italy}
\address{\ASU Arizona State University, Tempe, Arizona 85287}
\address{\SACLAY CEA-Saclay, Service de Physique Nucl\'eaire, F91191 Gif-sur-Yvette, Cedex, France}
\address{\UCLA University of California at Los Angeles, Los Angeles, California  90095}
\address{\CMU Carnegie Mellon University, Pittsburgh, Pennsylvania 15213}
\address{\CUA Catholic University of America, Washington, D.C. 20064}
\address{\CNU Christopher Newport University, Newport News, Virginia 23606}
\address{\UCONN University of Connecticut, Storrs, Connecticut 06269}
\address{\DUKE Duke University, Durham, North Carolina 27708}
\address{\EDINBURGH Edinburgh University, Edinburgh EH9 3JZ, United Kingdom}
\address{\FIU Florida International University, Miami, Florida 33199}
\address{\FSU Florida State University, Tallahassee, Florida 32306}
\address{\GWU The George Washington University, Washington, DC 20052}
\address{\GLASGOW University of Glasgow, Glasgow G12 8QQ, United Kingdom}
\address{\INFNGE Istituto Nazionale di Fisica Nucleare, Sezione di Genova, 16146 Genova, Italy}
\address{\ORSAY Institut de Physique Nucleaire ORSAY, IN2P3 BP 1, 91406 Orsay, France}
\address{\ITEP Institute of Theoretical and Experimental Physics, Moscow, 117259, Russia}
\address{\JMU James Madison University, Harrisonburg, Virginia 22807}
\address{\KYUNGPOOK Kyungpook National University, Taegu 702-701, South Korea}
\address{\MIT Massachusetts Institute of Technology, Cambridge, Massachusetts  02139}
\address{\UMASS University of Massachusetts, Amherst, Massachusetts  01003}
\address{\MSU Moscow State University, 119899 Moscow, Russia}
\address{\UNH University of New Hampshire, Durham, New Hampshire 03824}
\address{\NSU Norfolk State University, Norfolk, Virginia 23504}
\address{\OHIOU Ohio University, Athens, Ohio 45701}
\address{\ODU Old Dominion University, Norfolk, Virginia 23529}
\address{\PITT University of Pittsburgh, Pittsburgh, Pennsylvania 15260}
\address{\RPI Rensselaer Polytechnic Institute, Troy, New York 12180}
\address{\RICE Rice University, Houston, Texas 77005}
\address{\URICH University of Richmond, Richmond, Virginia 23173}
\address{\SCAROLINA University of South Carolina, Columbia, South Carolina 29208}
\address{\UTEP University of Texas at El Paso, El Paso, Texas 79968}
\address{\JLAB Thomas Jefferson National Accelerator Facility, Newport News, Virginia 23606}
\address{\UCONY Union College, Schenectady, New Yotk  12308}
\address{\VT Virginia Polytechnic Institute and State University, Blacksburg, Virginia   24061}
\address{\VIRGINIA University of Virginia, Charlottesville, Virginia 22901}
\address{\WM College of William and Mary, Williamsburg, Virginia 23187}
\address{\YEREVAN Yerevan Physics Institute, 375036 Yerevan, Armenia}
\date{\today}

\begin{abstract}
Nearly complete angular distributions of the two-body deuteron photodisintegration 
differential cross section have been measured using the CLAS detector and the tagged 
photon beam at JLab. The data cover photon energies between 0.5 and 3.0~GeV and center-of-mass 
proton scattering angles \mbox{$10^\circ$--$160^\circ$}.
The data show a persistent forward-backward angle asymmetry over the explored energy range, 
and are well-described by the non-perturbative Quark Gluon String Model.
\end{abstract}


\maketitle

\section{INTRODUCTION}

Quantum Chromodynamics (QCD) has been successfully applied in describing the structure and 
production of hadrons at high energies where perturbation theory can be used. There one can 
derive QCD scaling laws for the cross sections and hadronic helicity conservation laws. 
However, nuclear reactions have been conventionally described in terms of baryons and mesons 
rather than quarks and gluons. It is therefore interesting and important to know in which 
energy region the transition from hadronic picture to quark-gluon picture takes place. This 
is why major efforts in nuclear physics have been devoted, both 
theoretically and experimentally, to looking for qualitatively new phenomena which arise from 
the underlying quark degrees-of-freedom, and which cannot be modeled using meson field theories.\\
\indent
Deuteron photodisintegration at high energies is especially suited for this study, because a relatively 
large amount of momentum is transferred to the nucleons for a relatively low incident photon 
energy~\cite{GilGro,holt}.
One possible signature for the transition from nucleon-meson to quark-gluon degrees of freedom
is the scaling of reaction cross sections above some incident photon energy. 
In particular, simple Constituent Counting Rules (CCR) \cite{Matveev,Brodsky} predict an asymptotic 
${\it s^{-11}}$ dependence of $d\sigma/dt$ of the process at all proton angles. Here $s$ and $t$ are 
the invariant Mandelstam variables for the total energy squared and the four-momentum transfer squared, respectively.\\
\indent
Deuteron photodisintegration cross-sections above 1.2~GeV are available for photon energies 
$E_\gamma $ up to 5~GeV at three center-of-mass proton angles, 
$\vartheta_p^{\rm{c.m.}} =36^{\circ}, 52^{\circ}, 69^{\circ}$; up to 4~GeV at 
$\vartheta_p^{\rm{c.m.}} =90^{\circ}$~\cite{SLACNE81,SLACNE82,SLACNE17,Bochna,Schulte}; and at eight angles 
with $\vartheta_p^{\rm{c.m.}} = 30^{\circ}-143^{\circ}$, for $E_\gamma= $ 1.6, 1.9, and 
2.4~GeV~\cite{SchulteA}. 
The asymptotic scaling predicted by CCR is observed at $\vartheta_p^{\rm{c.m.}} =69^{\circ}$ and 
$90^{\circ}$ already at $E_\gamma =1$~GeV and at $\vartheta_p^{\rm{c.m.}} =52^{\circ}$ and 
$36^{\circ}$ only from 3 and 4~GeV, respectively. 
In contrast, polarization observables measured at $\vartheta_p^{\rm{c.m.}} =90^{\circ}$ 
for photon energies up to 2~GeV ~\cite{pol,adam} do not support hadronic helicity conservation 
predicted by pQCD. Thus, it seems that although the observation of the scaling in the cross section 
at a few proton angles indicates the onset of the quark-gluon degrees of freedom, the appropriate 
underlying physics has a mixture of perturbative and non-perturbative QCD aspects.\\
\indent
In this context, several non-pQCD models attempt to account for the experimental results using different 
strategies. The Reduced Nuclear Amplitude model (RNA)~\cite{rna} incorporates some of the soft 
physics not described by pQCD by using experimentally determined nucleon form factors to describe 
the gluon exchanges within the nucleons. The RNA calculation is only available at 
$\vartheta_p^{\rm{c.m.}} =90^{\circ}$ and makes no predictions for the angular dependence of the 
cross section. The calculations are normalized to data at energies sufficiently large, assuming that 
perturbative regime is reached.
\\
\indent
The Hard Quark Rescattering Mechanism model (HRM)~\cite{hrm,hrm-pre1} assumes 
that the photon is absorbed by a quark in one nucleon, followed by a high momentum transfer 
with a quark of the other nucleon leading to the production of two nucleons with high relative 
momentum. The nuclear scattering amplitude is expressed as a convolution of the large $pn$ 
scattering amplitude, the hard photon-quark interaction vertex and the low-momentum nuclear wave 
function. The authors use experimental data for the $pn$ cross section, but since data 
do not exist for the actual kinematic conditions needed, they must be extrapolated, and predictions 
for deuteron photodisintegration are given as a band corresponding to the uncertainties introduced 
by the extrapolations. The model provides a parameter-free prediction of $d\sigma/dt$ at 
$\vartheta_p^{\rm{c.m.}} =90^{\circ}$, and introduces a phenomenological function $f(t/s)$ that is 
close to unity at $\vartheta_p^{\rm{c.m.}} =90^{\circ}$, and varies slowly with 
$\vartheta_p^{\rm{c.m.}}$.
Another attempt ~\cite{hrm-diaz} to describe the deuteron photodisintegration within the same theoretical framework 
of HRM, using an exact calculation of the quark exchange amplitude, provides evidence that the assumption used in ~\cite{hrm,hrm-pre1} are questionable.
\\
\indent
The Quark Gluon String model (QGS)~\cite{leonid,gris,gris2} describes the reaction as proceeding 
through three-quark exchange, with an arbitrary number of gluon exchanges. The exchanged nucleon 
is replaced by a nucleon Regge trajectory that represents the sum of a tower of exchanged nucleon 
resonances. The best description of the data is obtained using a nonlinear Regge trajectory. The 
model takes all but two of its free parameters from other processes, and fixes the remaining two 
using the experimental data on the deuteron photodisintegration cross section at 
$E_\gamma = 1.6$~GeV and $\vartheta_p^{\rm{c.m.}} = 36^{\circ}$ and $52^{\circ}$. It provides the 
angular distributions and polarization observables for few-GeV beam energies. 
\\
\indent
Despite appearances, hard deuteron photodisintegration is an intractable problem in meson-baryon 
theories, the Asymptotic Meson Exchange Current model (AMEC)~\cite{amec} is able to extrapolate the conventional 
$N-\pi$ interaction mechanisms to higher energy using form factors to describe the $dNN$ interaction 
vertex, and fix an overall normalization factor by fitting the experimental data at 1~GeV. 
\\
\indent
A better insight into the 
competing models can be obtained from more detailed angular distributions of differential 
cross sections over broader angular and energy ranges than those presently available, and for final 
states involving different polarizations of the final hadrons.\\
\indent
We report here the first measurement of nearly-complete angular distributions 
(\mbox{$10^\circ \leq \vartheta_p^{\rm{c.m.}} \leq 160^\circ$}) of the two-body deuteron 
photodisintegration cross section obtained with the CEBAF Large Acceptance Spectrometer (CLAS) in 
Hall B at the Thomas Jefferson National Accelerator Facility (experiment E93-017)~\cite{Rossi} 
for photon energies between 0.5 and 3~GeV. The data offer the opportunity for a  
detailed study of the energy dependence of differential cross section of the reaction at fixed 
proton angles, aiming at determining the onset of asymptotic scaling~\cite{RossiPRL}.\\
\indent
In the following, we first give some details of the experiment (Sec.~II) and its data 
analysis (Sec.~III). 
Then, we present our results on the deuteron photodisintegration cross sections $d\sigma / d\Omega$ 
and ${d\sigma}/{dt}$, and compare them to available theoretical models and existing data (Sec.~IV). 
We conclude with a summary (Sec.~V).

\section{EXPERIMENTAL SETUP}
The data described in this paper were collected at the Thomas Jefferson National Accelerator Facility 
(JLab) during a 32-day run in August and September 1999 using the Hall~B tagged photon beam~\cite{tagging} 
and the CLAS detector~\cite{mecking}.
The {bremsstrahlung} photon beam was produced by a \mbox{$10-13$~nA} continuous electron beam of energy 
$E_0=$~2.5~GeV (August) and 3.1~GeV (September) impinging on a gold foil of $10^{-4}$ radiation lengths. 
A tagging spectrometer, with an energy resolution of $0.1E_0~\%$, was used to tag $\sim 10^7$ photons per 
second in the energy range \mbox{$(0.20-0.95)E_0$}.\\
\indent
A cylindrical mylar cryogenic target 10~cm long and 4~cm in diameter was filled with liquid deuterium at 
about 23.7 K. 
The final-state particles were detected in the CLAS spectrometer, which is built around six superconducting 
coils producing a toroidal magnetic field symmetric about the beam and oriented primarily in the azimuthal 
direction. 
The coils naturally separate the detector into six sectors, each functioning as an independent magnetic 
spectrometer.  
Each sector is instrumented with 3 sets of multi-wire drift chambers for track reconstruction and one layer 
of scintillator counters, covering the angular range from $8^\circ$ to $143^\circ$, for time-of-flight 
measurements. The forward region (\mbox{$8^\circ \leq \vartheta \leq 45^\circ$}) contains gas-filled threshold 
Cherenkov counters and lead-scintillator sandwich-type electromagnetic calorimeters for particle identification.
For two CLAS sectors the coverage of the electromagnetic calorimeters is extended up to polar angles of 
$70^\circ$. 
The trigger for the data acquisition was defined by the coincidence between a signal in the tagger (identified 
photon) and one charged hadron in CLAS. 
Under these conditions 1771~million events were collected. 

\section{DATA ANALYSIS}
\subsection{Data selection}
\indent
A data quality check was performed to select runs with stable beam and detector performance. First, several 
run-based parameters normalized to the incident photon flux were required to be constant at the few percent 
level from run to run: {\it a)} the total number of charged particles, and {\it b)} the number of particles 
$p$, $\pi^+$, $\pi^-$, $K^{+}$, and $K^{-}$. Then, {\it c)} the number of triggers with at least one charged 
particle in the final state for each tagger-timing counter; {\it d)} the number of photodisintegration events 
per 100~MeV; and {\it e)} the number of photodisintegration events per CLAS 
sector were required to be stable within the statistical errors.\\
\indent
After applying the above data quality criteria, about 7\% of the originally collected data had been discarded.

\subsection{Event selection}
\indent
Photodisintegration events \mbox{$\gamma d \rightarrow pn $} were identified as follows:

\begin{itemize}
\item The software coincidence time-window between the tagger and CLAS was set to \mbox{$\pm 1$~ns}, since 
the machine electron bunches are separated by 2.004~ns. \item Only events with a single charged particle, 
the proton, in the final state were selected. Protons were identified by determining momentum and path length 
using the drift chambers, and velocity from the time-of-flight counters. 
\item The reconstructed vertex position of the proton along with beam line was used to remove events 
originating outside the target cell.
\item Cuts on the square of the missing mass \mbox{$M_X^2 = (P_\gamma + P_d - P_p)^ 2$} were performed to 
select exclusive two-body deuteron photodisintegration events. 
Here $P_\gamma$, $P_d$, and $P_p$ are the \mbox{four-momenta} of the photon, deuteron, and proton, respectively. 
In this study $M_X$ is the mass of the neutron.
\end{itemize}

\subsection{Momentum correction}
The momentum of the charged particles measured with CLAS strongly relies on the correct knowledge of the magnetic 
field geometry and the positioning of the drift chambers. 
Due to the complexity of the detector, and particularly of the superconducting torus magnet assembly, it is 
crucial to make sure that the momentum determined by the drift chamber tracking system is reliable. 
For this reason the position of the peak of the missing mass distributions from \mbox{$\gamma d \rightarrow p X$} 
events has been checked over the whole range of proton momenta and scattering angles. 
After correcting for the energy loss in the target, the value of the peak was slightly off with respect to the 
neutron rest mass depending on the proton scattering angle.
\\
\indent
A correction procedure has been applied to the data using  an empirical function depending only on the measured 
three-momentum of the proton. It was assumed that the proton track angles are correctly measured, since the CLAS 
angular resolution is much better than the momentum resolution~\cite{mecking}. 
We have also checked that the contribution due to the photon energy uncertainty is negligible, by using exclusive 
\mbox{$\gamma d \rightarrow pp\pi^-$} events.
The correction function has been calculated for each kinematic bin by fitting the ratio of the expected momentum, 
as calculated from the photon energy and the proton scattering angle, to the measured momentum~\cite{Mirazita}.\\
\indent
The correction procedure introduced a significant improvement in the resulting width and position of the peak in 
the missing mass distributions. 
\begin{figure}[htbp]
\begin{center}
\includegraphics[width=8.0cm, height=7.0cm]{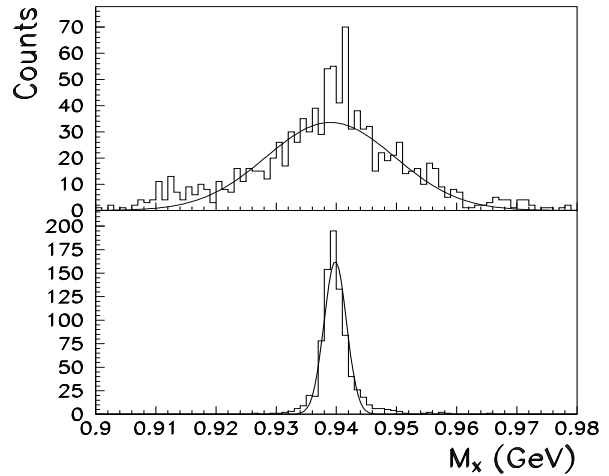}
\caption{\small Distributions for the $M_X$ peak values for \mbox{$\gamma d \rightarrow pX$} events before (top) 
and after (bottom) the momentum corrections are applied. The width of the corrected distribution is smaller (13.2~MeV rms $\rightarrow$ 
3.2~MeV rms) and the centroid is closer to the neutron rest mass.}
\label{fig:pcorr}
\end{center}
\end{figure}
Figure~\ref{fig:pcorr} shows the distributions for the peak values of the $M_X$ distributions for 
\mbox{$\gamma d \rightarrow pX$} events for all CLAS sectors and all runs, before and after the correction was 
applied. Clearly, after the correction the distribution of the peak values is sharper (13.2~MeV rms $\rightarrow$ 
3.2~MeV rms), and its mean value is closer to the neutron rest mass. 

\subsection{Efficiency}
\begin{figure}[h]
\begin{center}
\hspace*{-1.0cm}
\includegraphics[width=9.cm, height=6.cm]{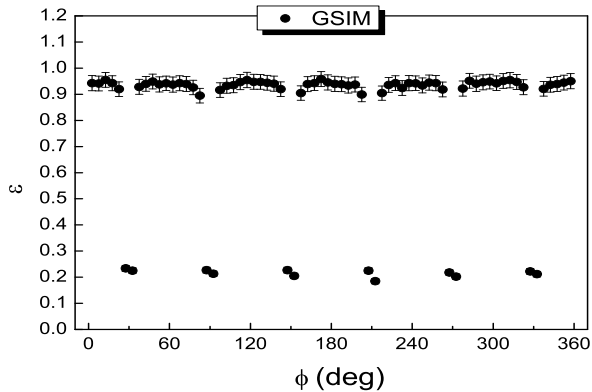}
\caption{\small  The CLAS proton detection efficiency evaluated using GSIM as a function of the azimuthal angle 
$\varphi$ for $P_p^{\rm{LAB}}$=0.95~GeV, $\vartheta_p^{\rm{LAB}}=65^\circ$, and $\varphi$ bins of $5^\circ$. 
}
\label{fig:gsim-5deg}
\end{center}
\end{figure}

The single-proton detection efficiency in CLAS cannot be extracted from the deuteron photodisintegration data 
itself over the whole kinematic region of emitted protons. Exclusive events, where both the neutron and the proton 
are detected, are limited because neutrons could be detected in the calorimeters only over a small angular range 
\mbox{$\vartheta^{\rm{LAB}} \leq 45^\circ$} for four CLAS sectors and 
\mbox{$\vartheta^{\rm{LAB}} \leq 70^\circ$} for the other two sectors. 
Other reaction channels with additional particles are problematic because the photodisintegration protons have the 
highest momentum for a given proton angle and photon energy.\\
\indent
For these reasons, the single proton efficiency has been evaluated using a GEANT simulation (GSIM) of the CLAS 
detector. Photodisintegration events have been generated uniformly in proton momentum and angle in the laboratory 
system, and then have been analyzed following the standard reconstruction chain~\cite{Mirazita}.\\
\indent
The proton detection efficiency $\epsilon_{\rm{GSIM}}$ has been calculated in each kinematic bin in the laboratory 
system as the ratio of reconstructed protons $N_{\rm{REC}}$ to generated ones $N_{\rm{GEN}}$:
\begin{equation}
\epsilon_{\rm{GSIM}}  = \frac {N_{\rm{REC}}} {N_{\rm{GEN}}} \ .
\label{eq:gsim-epsilon}
\end{equation}\\
Bin widths for proton momentum  \mbox{$\Delta P_p^{\rm{LAB}}=0.1$~GeV/c} and polar scattering angle  
$\Delta \vartheta_p^{\rm{LAB}} =10^\circ$ have been chosen. A smaller azimuthal angle bin of width 
$\Delta \varphi_p^{\rm{LAB}}=5^\circ$ has been selected to better investigate the azimuthal behavior of the CLAS 
proton detection efficiency, which gets worse on the boundaries of each sector due to the presence of the magnet 
coils.\\
\indent
As an example, Fig.~\ref{fig:gsim-5deg} shows the resulting proton detection efficiency for  \mbox{$P_p^{\rm{LAB}}$
= 0.95~GeV} and $\vartheta_p^{\rm{LAB}}=65^\circ$ as a function of the azimuthal angle $\varphi_p^{\rm{LAB}}$. 
Similar plots have been obtained for the other proton angles. 
For $\vartheta_p^{\rm{LAB}}=45^\circ-125^\circ$, the proton efficiency is nearly constant in the central region 
of each sector, with an average value of $(94 \pm 1)\%$, and drops sharply near the sector boundaries. At forward 
angles, the average efficiency decreases dropping to about $(50 \pm 1)\%$ at $\vartheta_p^{\rm{LAB}}=15^\circ$.

\begin{figure}[htbp]
\begin{center}
\hspace*{-1.0cm}
\includegraphics[width=9.cm, height=6.cm]{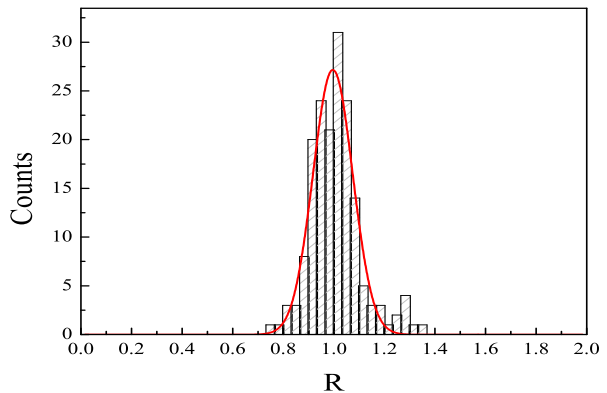}
  \caption{\small Distribution of the ratio \mbox{$R = \epsilon_{\rm{DATA}}/ \epsilon_{\rm{GSIM}}$} between the 
  proton detection efficiency measured from the data using the \mbox{$\gamma d \rightarrow p \pi^- p$} reaction  
  and that obtained from GSIM, for proton momenta in the interval \mbox{$0.5-1.1$~GeV} and for the central 
  $20^\circ$ in $\varphi$ for all sectors.}
\label{fig:ratio}
\end{center}
\end{figure}

In order to check the reliability of the simulations, the proton detection efficiency has also been obtained using 
the data (where they are available) from the overdetermined \mbox{$ \gamma d \rightarrow p p \pi^{-}$} reaction. 
Each time a $p$ and $\pi^-$ pair is found, the missing mass is calculated and the three-momentum is computed for 
candidates within tight constraints on the proton missing mass. The ratio between the number of exclusive 
\mbox{$p \pi^- p$} events, $N^{p \pi^- p }$, found by the particle identification, and the number of events with 
$X$ identified as a proton by the missing mass cuts, $N^{p \pi^- X(p)}$, minus the number of the background 
events under the missing mass peak of the $p \pi^- X$ distribution, $N_B$, gives the experimental detection efficiency: 
\begin{equation}
\epsilon_{\rm{DATA}} 
= \frac{N^{p \pi^- p }} 
{N^{p \pi^- X(p) } - N_B} \ .
\label{eq:data-epsilon}
\end{equation}
The distribution of the values of the ratio \mbox{$R = \epsilon_{\rm{DATA}} / \epsilon_{\rm{GSIM}}$} calculated 
where both $\epsilon_{\rm{DATA}}$ and $\epsilon_{\rm{GSIM}}$ are available with good statistics (proton momenta 
in the range \mbox{$0.5-1.1$~GeV} and central regions of the six CLAS sectors) is shown in Fig.~\ref{fig:ratio}. 
The mean value of the distribution is very close to unity ($0.997 \pm 0.003$).\\
\indent
We checked also that regions of lower efficiency in CLAS corresponding to dead time-of-flight paddles or drift 
chambers wires are well-reproduced by the simulation. Thus, the comparison over limited kinematics validates the 
GSIM results.\\
\subsection{Fiducial cuts and mean efficiencies}
As shown in Fig.~\ref{fig:gsim-5deg} the proton detection efficiency is constant in the central azimuthal regions 
of the six CLAS sectors and decreases steeply near the sector boundaries. 
Thus, only events in a fiducial region ({\em i.e.}~azimuthal region of the phase space where the efficiency is uniform) 
of the detector have been used. 
For each bin in proton momentum and scattering angle, and for each CLAS sector S, a mean efficiency $\epsilon$ 
is defined as:
\begin{gather}
\epsilon \left( \Delta P_p^{\rm{LAB}},\Delta \vartheta_p^{\rm{LAB}}, S \right) =  \\
\langle\epsilon_{\rm{GSIM}} \left ( \Delta P_p^{\rm{LAB}},\Delta \vartheta_p^{\rm{LAB}}, S \right)\rangle 
\cdot \eta \left ( \Delta P_p^{\rm{LAB}},\Delta \vartheta_p^{\rm{LAB}}, S \right)\ . \notag
\label{eq:gsim-sector}
\end{gather}
in which 
$\langle\epsilon_{\rm{GSIM}} \left ( \Delta P_p^{\rm{LAB}},\Delta \vartheta_p^{\rm{LAB}}, S \right)\rangle$ 
is the average proton detection efficiency over the fiducial $\Delta\varphi$ region, 
and $\eta \left ( \Delta P_p^{\rm{LAB}},\Delta \vartheta_p^{\rm{LAB}}, S \right)$ is the portion 
of the CLAS sector inside the fiducial cuts ({\em i.e.}~the fraction of the $\varphi$ interval 
considered). 
\subsection{Background subtraction}
\begin{figure}[ht]
\begin{center}
\includegraphics[width=8.cm, height=6.cm]{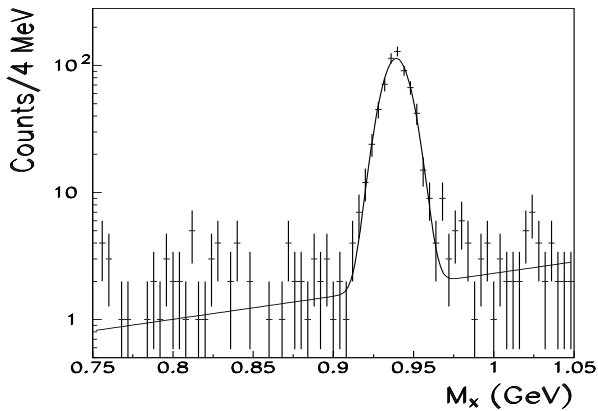}
\caption{\small Typical missing-mass spectrum of the reaction \mbox{$\gamma d \rightarrow p X$} obtained in Sector 
5 for photon energy of $E_{\gamma}=$0.95~GeV and proton scattering angle $\vartheta_p^{\rm{LAB}}=25^\circ$.
}
\label{fig:id1}
\end{center}
\end{figure}

\begin{figure}[htbp]
\begin{center}
\includegraphics[width=8.cm, height=7.cm]{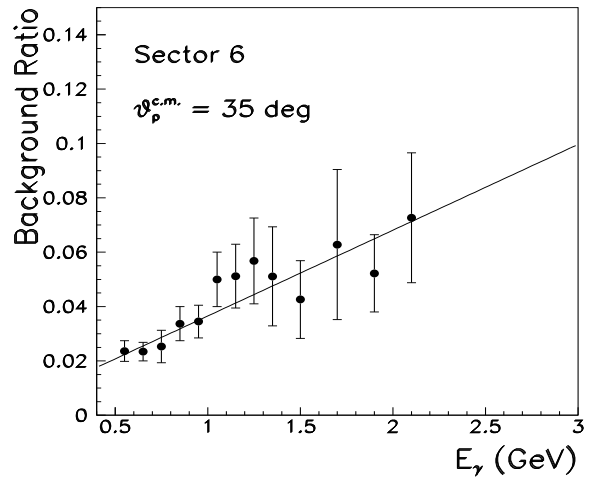}
\caption{\small The behavior of the background contribution $k$ as a function of the photon energy for proton 
scattering angles \mbox{$30^\circ \leq \vartheta_p^{\rm{c.m.}} \le 40^\circ$} and CLAS Sector 6.}
\label{fig:bkg1}
\end{center}
\end{figure}

At all proton angles and photon energies, the missing-mass distributions of the \mbox{$\gamma d \rightarrow p X $} 
reaction show a neutron mass peak $M_n$ riding on  a smooth background. As an example, figure~\ref{fig:id1} shows 
the missing-mass distribution obtained for incident photon energy $E_\gamma=$~0.95~GeV and proton scattering angle 
$\vartheta_p^{\rm{LAB}}=25^\circ$. The logarithmic scale emphasizes the background contribution. 
These distributions are well-reproduced with a Gaussian plus exponential form. Events within $\pm3 \sigma$ of the 
neutron peak have been kept for the determination of the cross section (here $\sigma$ is the width of the Gaussian 
distribution).\\
\indent
The background contribution $N_B$ to the number of total events under the peak $N_{\rm{peak}}$ has been evaluated 
by integrating the exponential fit function between the missing-mass cuts.
At photon energies higher than $(2.0-2.4)$~GeV depending on the proton angle, the neutron mass peaks are less clearly 
identifiable due to the low statistics. 
In these cases, the missing mass cuts for the selection of exclusive events have been obtained using a second-order 
polynomial fit in $\vartheta_p^{\rm{c.m.}}$ and $E_\gamma$, of the $(M_{\rm{peak}} + 3 \sigma)$ and  
$(M_{\rm{peak}} - 3 \sigma)$ values determined at lower photon energies.
In each $\vartheta_p^{\rm{c.m.}}$ bin the background has been evaluated by using a linear extrapolation of the fits 
in $E_{\gamma}$ to the ratio $k = N_B/N_{\rm{peak}}$ obtained at lower photon energies.
As an example, Fig.~\ref{fig:bkg1} shows the values of the ratios $k$ obtained for 
\mbox{$30^\circ \leq \vartheta_p^{\rm{c.m.}} \le 40^\circ$} for CLAS Sector 6. Similar plots are obtained for other 
proton angles and CLAS sectors. Here $k$ increases with the photon energy and proton angle $\vartheta_p^{\rm{c.m.}}$ 
due to the loss in momentum resolution.\\
\indent
In order to check the extrapolation procedure in the photon energy region above $(2.0-2.4)$~GeV, the background 
contribution has been evaluated from the data using larger bins (then increasing the statistics and making clearly 
identifiable the peaks) and compared to the result obtained from the extrapolation. The values have been found to 
be in a very good agreement with each other.

\subsection{Photon Flux}
\begin{figure}[h]
 \begin{center}
 \includegraphics[width=8.cm, height=7.cm]{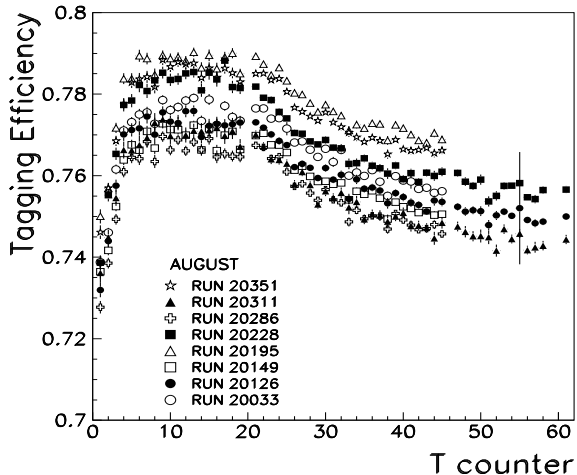}
 \caption{\small Tagging efficiency for the 61 tagger-timing counters, measured during the $E_0 = 2.5$~GeV normalization 
runs. In some runs (open points) the low energy counters had been switched off to increase the statistics at high 
photon energies.
}
\label{fig:tag-eff}
 \end{center}
\end{figure}

The incident flux of photons on the target is given by $N_\gamma= N_{e} \cdot \epsilon_{T}$, where $N_{e}$ is the 
number of tagged electrons, and $\epsilon_{T}$ is the tagging efficiency. $N_{e}$ has been measured online during 
the production runs, while $\epsilon_{T}$ has been measured during the normalization runs at low intensity 
($\sim 10^5$$\gamma/s$) using a nearly $100\%$ efficiency lead-glass total absorption counter. We assume that 
$\epsilon_{T}$ remains unchanged during the production runs. 
Normalization runs were performed every time the experimental conditions for production runs were changed: a total 
of 178~million normalization events were collected.
Fig.~\ref{fig:tag-eff} shows the tagging efficiency measured in all the normalization 
runs at $E_0$=2.5~GeV. Similar results were obtained in the normalization runs at  $E_0$=3.1~GeV.
The tagging efficiency is stable at a level of $\simeq 2\%$. 

\subsection{Systematic Uncertainties}
The contributions to the overall systematic uncertainty come from \emph{i)} the determination of the number of 
incident photons, $\approx 1.9 \%$, evaluated by looking at the variation of the number of photons per tagger channel 
in normalization runs; \emph{ii)} the determination of the target length and density, $\approx 0.5 \%$; \emph{iii)} 
the proton detection efficiency, $(2-8)\%$, evaluated as 
\mbox{$(\epsilon_{\rm{GSIM}}- \epsilon_{\rm{DATA}})/\epsilon_{\rm{GSIM}}$}; and \emph{iv)} the background subtraction, 
around $(1-2)\%$ for photon energies below 1~GeV and higher (up to $\sim 6\%$) at forward and backward angles where 
the detector resolution and efficiency are worse. 
The latter has been evaluated by repeating the data analysis using both missing-mass cuts reduced and enlarged by 
$20\%$, and looking at the variation of the differential cross section.\\
\indent
The resulting total systematic error is $\leq 10 \%$ in the whole measured range.

\section{RESULTS}
\begin{figure*}[htbp]
\begin{center}
\vspace*{-1.0cm} 
\includegraphics[width=17.cm, height=23.cm]{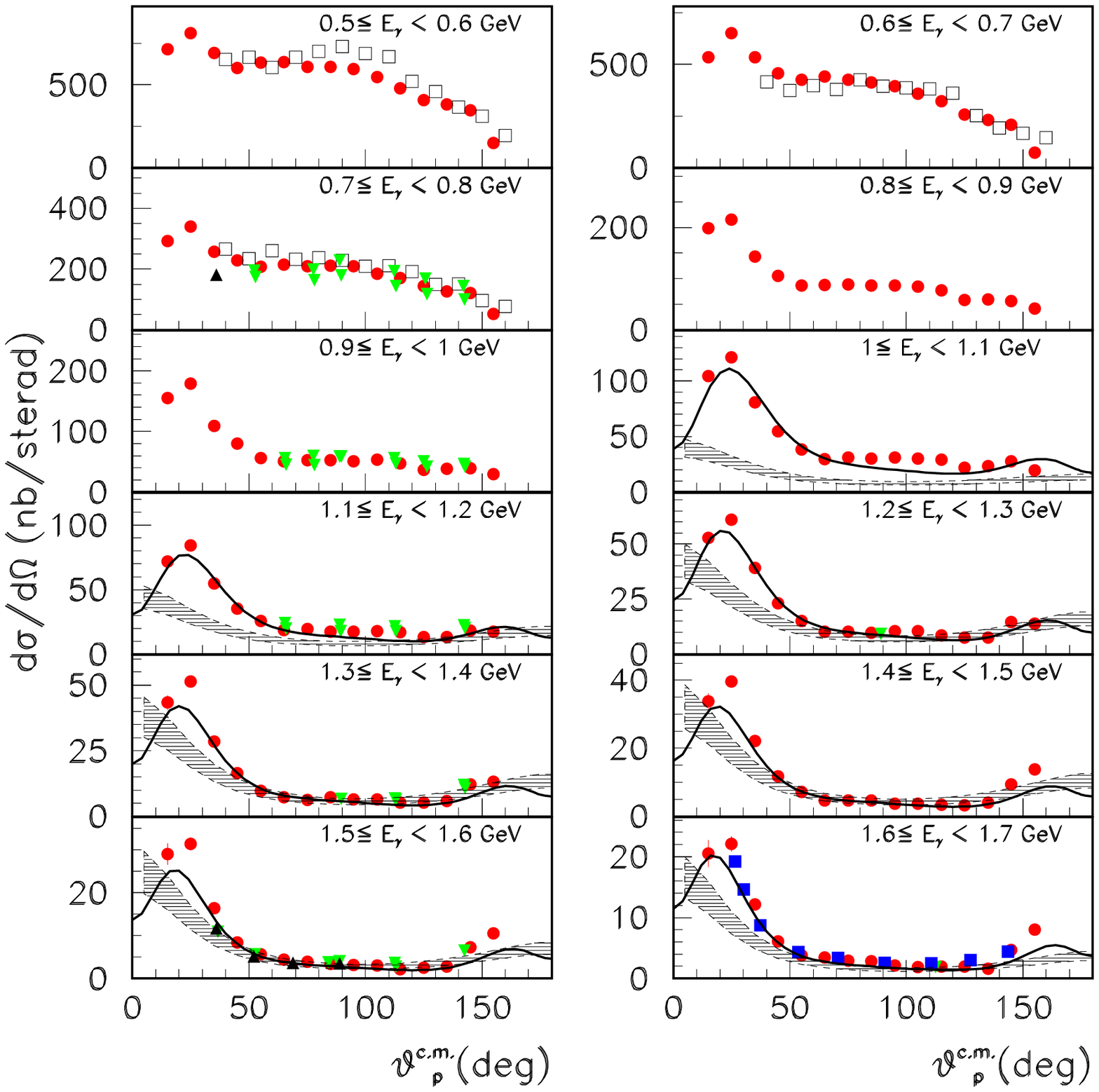}
\caption{\small (Color) Angular distributions of the deuteron photodisintegration cross section measured by the CLAS (full/red 
circles) in the incident photon energy range \mbox{$0.50-1.70$~GeV}.
Results from Mainz~\cite{Mainz} (open squares, average of the measured values in the given photon energy intervals), 
SLAC~\cite{SLACNE17,SLACNE81,SLACNE82} (full/green down-triangles), JLab Hall~A~\cite{SchulteA} (full/blue squares) 
and Hall~C~\cite{Bochna,Schulte} (full/black up-triangles) are also shown. Error bars represent the statistical 
uncertainties only. The solid line and the hatched area represent the predictions of the QGS~\cite{gris} and the 
HRM~\cite{hrm-boh} models, respectively. 
}
\label{fig:dist1t}
\end{center}
\end{figure*}

\begin{figure*}[htbp]
\begin{center}
\vspace*{-1.0cm} 
\includegraphics[width=17.cm, height=23.cm]{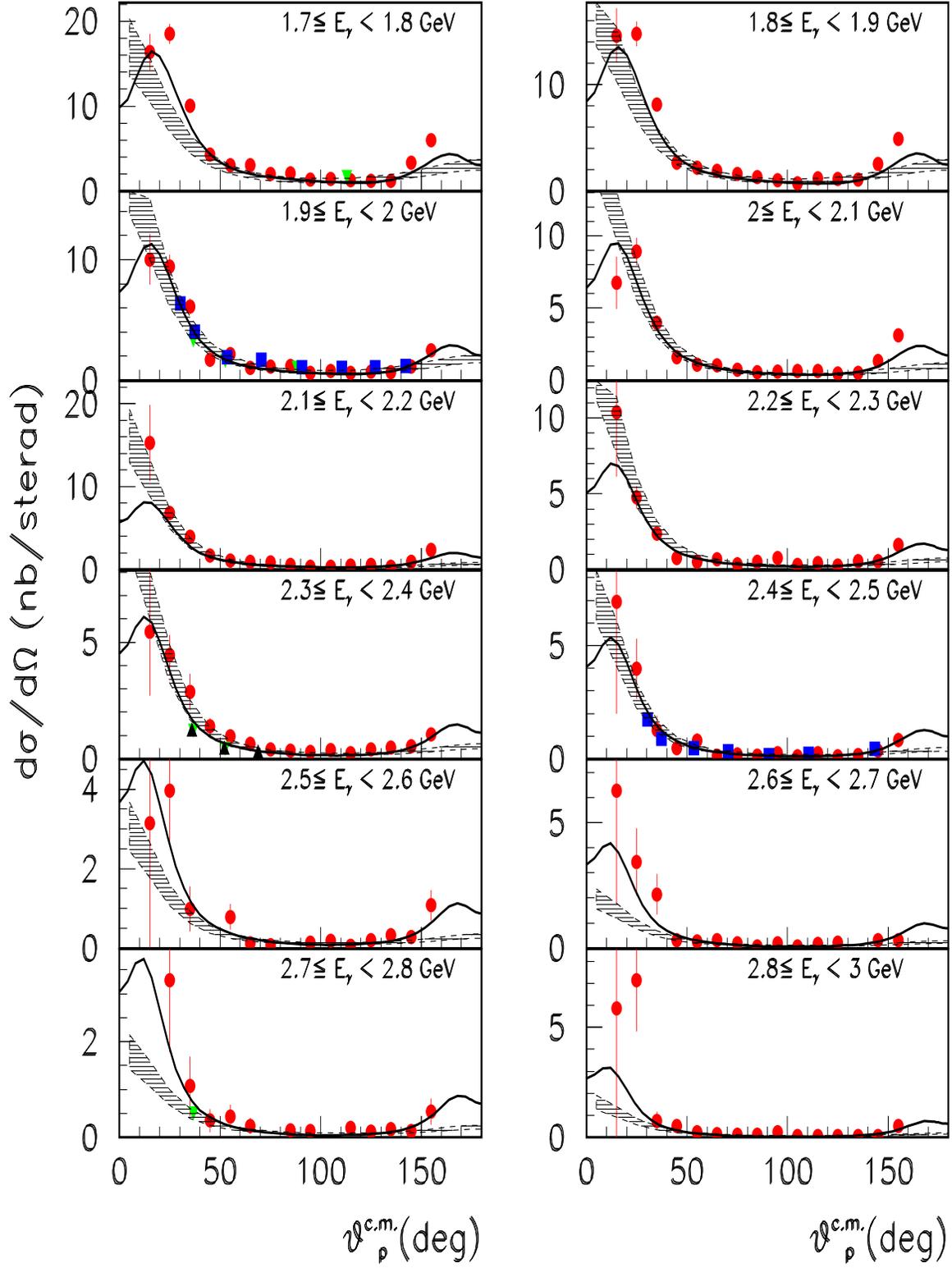}
 \caption{\small (Color) Same as Fig.~\ref{fig:dist1t} for photon energies \mbox{$1.7-3.0$~GeV}.
}
\label{fig:dist2t}
\end{center}
\end{figure*}

\begin{figure*}[htbp]
\begin{center}
\includegraphics[width=16.cm, height=16.cm]{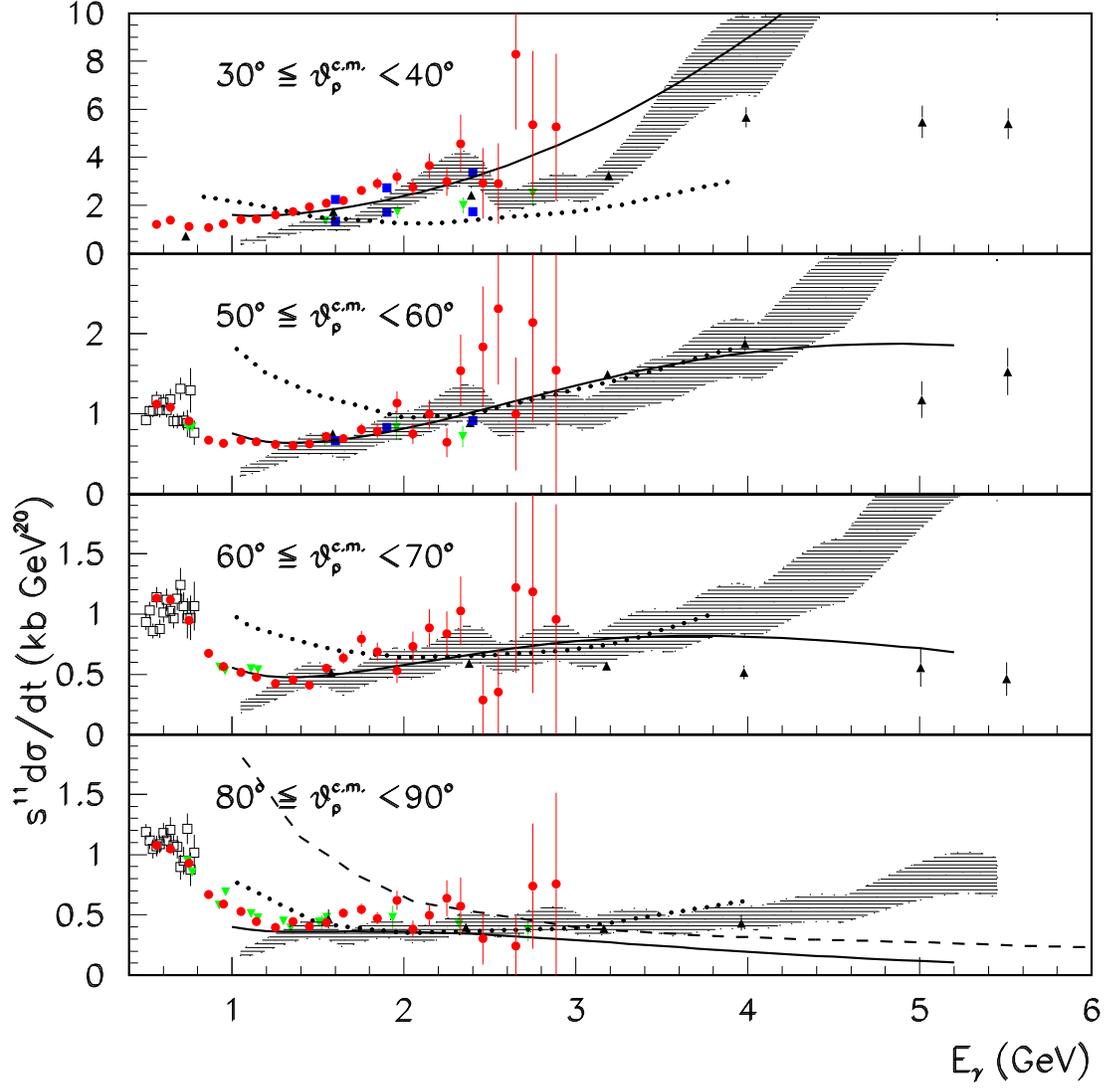}
\caption{\small 
(Color) Deuteron photodisintegration cross sections ${s^{11}d\sigma}/{dt}$ as a function of $E\gamma$ for the given proton 
scattering angles. Results from CLAS (full/red circles), Mainz~\cite{Mainz} (open squares), 
SLAC~\cite{SLACNE17,SLACNE81,SLACNE82} (full/green down-triangles), JLab Hall A~\cite{SchulteA} (full/blue 
squares) and Hall C~\cite{Bochna,Schulte} (full/black up-triangles) are included, as well as predictions 
of the QGS~\cite{gris} (solid line), AMEC~\cite{amec} and RNA~\cite{rna} models (dotted and dashed 
lines, respectively), and the HRM~\cite{hrm-boh} model (hatched area). Error bars represent the statistical 
uncertainties only.
}
\label{fig:fit1}
\end{center}
\end{figure*}

The photodisintegration cross section was calculated using:
\begin{gather}
\frac{d\sigma}{d\Omega}(E_\gamma,\vartheta_p^{\rm{c.m.}})= \\
\frac {A}{\rho x N_A}
\frac{N^W_{\rm{peak}}(E_\gamma, \vartheta_p^{\rm{c.m.}}) } 
{N_\gamma (E_\gamma)\, \Delta \Omega } 
\left( 1 - k(E_\gamma, \vartheta_p^{\rm{c.m.}}) \right) \notag , 
\label{eq:cross}
\end{gather}
where $N^W_{\rm{peak}}$ is the number of \mbox{$\gamma d \rightarrow pn$} events weighted by the efficiency, 
$\Delta\Omega$ is the solid angle, $A$ is the target molecular weight, $N_A$ is Avogadro's number, $\rho$ is 
the target density and $x$ the target effective length.\\
\indent
In Figures~\ref{fig:dist1t}~and~\ref{fig:dist2t} the angular distributions $d\sigma/d\Omega$ are shown as a 
function of $\vartheta_p^{\rm{c.m.}}$ for photon energy bins 100~MeV wide, in the range from 0.5 up to 3.0~GeV, and 
proton scattering angle bins $10^\circ$ wide in the range \mbox{$10^\circ \leq \vartheta_p^{\rm{c.m.}} \leq 160^\circ$}. 
The data are also given in Tables~\ref{tab:data2} and~\ref{tab:data4}. They are averaged over the six CLAS sectors.
The results obtained by the six CLAS sectors separately are consistent with each other within the systematic errors.\\
\indent 
This is the first measurement of the nearly complete angular distributions of the \mbox{$\gamma d \rightarrow pn$} 
reaction for photon energies between 0.5 and 3.0~GeV. It allows one to investigate the behavior of the cross section 
in the very forward and backward angular regions. The data show a clear forward/backward angle asymmetry in the whole 
range of explored photon energies. At high energies the cross sections increase at very forward and backward angles.\\
\indent
Also shown in Figures~\ref{fig:dist1t}~and~\ref{fig:dist2t} are the previous data and the predictions of the few 
available models.
For \mbox{$E_\gamma = 0.5-0.6$~GeV}, the Mainz data~\cite{Mainz} are slightly higher than CLAS at intermediate 
scattering angles.
Starting from \mbox{$E_\gamma = 0.7-0.8$~GeV} the comparison can be extended also to the 
SLAC~\cite{SLACNE17,SLACNE81,SLACNE82} and JLab Hall C~\cite{Bochna} data. The CLAS results agree well with the data 
from these experiments. 
For \mbox{$E_\gamma =1.6-1.7$~GeV}, \mbox{$E_\gamma =1.9-2.0$~GeV}, and \mbox{$E_\gamma =2.4-2.5$~GeV}, the CLAS 
results agree with the angular distributions measured by the JLab~Hall~A  collaboration~\cite{SchulteA} (the latter 
cover a smaller range in the proton scattering angle: \mbox{$26^\circ \leq \vartheta_p^{\rm{c.m.}} \leq 143^\circ$}), 
and extend to the very forward and backward angular regions where the cross section increases.\\ 
\indent
For $E_\gamma \geq 1.0$~GeV,  the predictions of the QGS model~\cite{gris,gris2} are shown in 
Figures~\ref{fig:dist1t}~and~\ref{fig:dist2t} (solid curve). This model describes the angular distributions very well, 
and accounts for the persistent forward/backward angle asymmetry seen in the data by invoking the interference of 
the isovector and isoscalar components of the photon. The interference is constructive at forward angles and destructive 
in the backward direction. The most forward points support the presence of the local maximum at about $20^\circ$ 
predicted by the model. 
The backward points do not extend far enough to check for the presence of the second maximum.
Also shown in Figures~\ref{fig:dist1t}~and~\ref{fig:dist2t} are the predictions of the HRM model~\cite{hrm-boh}
(hatched band) calculated using the best angular fit for fixed energy $pn$ scattering data. The bands reflect the 
poor accuracy of the data. The model agrees reasonably well with data in the central 
angular region over the whole explored energy range, and is lower at forward and backward angles apart from $E_\gamma= 1.8-2.5$~GeV. This agreement suggests a further investigation as, in principle, the HRM model is applicable for energies greater than $ \sim 2$~GeV.
\\
\indent
The rich amount of CLAS data has made a detailed study of the power law dependence $s^{-n}$ of the 
differential cross section $d\sigma/dt$ possible, in order to determine the onset threshold for the 
appearance of the $s^{-11}$ scaling law predicted by perturbative QCD. This study~\cite{RossiPRL} indicates 
a proton transverse momentum scaling thresholds of $P_T=1.0-1.3$~GeV/c for angles between $60^{\circ}$ 
and $130^{\circ}$, and $0.6-0.9$~GeV/c for forward and backward angles, with a nearly symmetric behavior 
around $90^{\circ}$.\\
\indent
Figure~\ref{fig:fit1} shows the results of $d\sigma/dt$ (full circles) multiplied by the factor $s^{11}$
predicted by CCR and plotted as a function of $E_\gamma$ for the four proton scattering angles for which the 
predictions from all existing models are available. Also shown in the figure are the previous data: Mainz~\cite{Mainz} (open squares), 
SLAC~\cite{SLACNE17,SLACNE81,SLACNE82} (full/green down-triangles), JLab Hall A~\cite{SchulteA} (full/blue 
squares) and Hall C~\cite{Bochna,Schulte} (full/black up-triangles). The two points at the same energy value from ~\cite{SchulteA} shown in the top panel come from two slightly different proton angles ($30.3 ^\circ$ and $37.4 ^\circ$).
The HRM model~\cite{hrm-boh} (hatched band) agrees reasonably well with data up to about 4~GeV, then tends 
to be higher at forward angles. 
The RNA calculation is only available at $\vartheta_p^{\rm{c.m.}} =90^{\circ}$. The estimate for this 
figure~\cite{bfgh} (dashed lines) is normalized to the datum at $E_\gamma= 3.16$~GeV. Other estimates at different 
angles have been provided in other papers by different authors but suffer from an incorrect normalization~\cite{hiller},
and therefore are not shown in the figure.
The AMEC model~\cite{amec} (dotted lines) predicts a slightly different energy dependence. The data at forward 
angles suggest a slower decrease of the cross section with energy than predicted. Moreover there is a 
discrepancy for the highest energy at $60^{\circ}-70^{\circ}$. Surprisingly, the model strongly overestimate 
data at energies lower than $1.6-2.0$~GeV.
The QGS model describes well the data at all four proton angles. The largest discrepancy is found at 
$30^{\circ}-40^{\circ}$ above 3~GeV where it suggests a slower decrease of the cross section with energy than observed.\\
\indent
Clearly, further theoretical developments in this non-perturbative regime would be desirable to understand 
the transition region between the meson exchange picture and the QCD description of high energy nuclear reactions.

\section{SUMMARY}
Differential cross sections for \mbox{$\gamma d \rightarrow pn$} have been measured for the first time with a nearly 
complete angular coverage (\mbox{$10^\circ \leq \vartheta_p^{\rm{c.m.}} \leq 160^\circ$}) in the photon energy range 
from 0.5 to 3.0~GeV using the CLAS detector and the tagged photon beam of Hall B at Jefferson Lab.
The shapes of the angular distributions $d\sigma/d\Omega$ show a persistent forward/backward angle asymmetry 
over the whole explored energy range.
The cross sections $d\sigma/dt$ fall by 2--3 orders of magnitude from 1 to 3 GeV photon energy. The data have 
been used to determine the scaling threshold at every proton angle between $10^\circ$ and $150^\circ$.
The non-perturbative Hard Quark Rescattering Mechanism and Quark Gluon String models describe the data well. The latter accounts well also for the forward and 
backward angle asymmetry.

\section{ACKNOWLEDGMENTS}
We would like to acknowledge the outstanding efforts of the staff of the Accelerator and the Physics Divisions at 
JLab that made this experiment possible. This work was supported in part by  the  Italian Istituto Nazionale di Fisica 
Nucleare, the French Centre National de la Recherche Scientifique and the Commissariat \`{a} l'Energie Atomique, 
the U.S. Department of Energy and the National Science Foundation, and the Korea Science and Engineering Foundation. 
The Southeastern Universities Research Association (SURA) operates the Thomas Jefferson National Accelerator Facility 
for the United States Department of Energy under contract DE-AC05-84ER40150.

\newpage

\newpage

\begin{table*}[htbp]
 \begin{center}
 \begin{tabular}{|c|ccccc|ccccc|ccccc|ccccc|} \hline\hline  
  &\multicolumn{5}{c|}{}  
  &\multicolumn{5}{c|}{}  
  &\multicolumn{5}{c|}{}  
  &\multicolumn{5}{c|}{}
 \\                              
  $\langle E_\gamma \rangle $ (GeV) 
  &\multicolumn{5}{c|}{$10^\circ\leq \theta_p^{\rm{c.m.}} <20^\circ$} 
  &\multicolumn{5}{c|}{$20^\circ\leq \theta_p^{\rm{c.m.}} <30^\circ$}
  &\multicolumn{5}{c|}{$30^\circ\leq \theta_p^{\rm{c.m.}} <40^\circ$} 
  &\multicolumn{5}{c|}{$40^\circ\leq \theta_p^{\rm{c.m.}} <50$} 
\\
  &\multicolumn{5}{c|}{}  
  &\multicolumn{5}{c|}{}  
  &\multicolumn{5}{c|}{}  
  &\multicolumn{5}{c|}{}\\
\hline
 0.560 &  714.5 & $\pm$ & 16.9 & $\pm$ & 40.2 &  811.2 & $\pm$ &  11.6 & $\pm$ & 48.3 &  693.2 & $\pm$ & 8.6   & $\pm$ & 58.3  &  601.1 & $\pm$ & 7.0   & $\pm$ & 41.3  \\
 0.641 &  534.3 & $\pm$ & 11.6 & $\pm$ & 30.3 &  651.0 & $\pm$ &  8.4  & $\pm$ & 38.8 &  533.8 & $\pm$ & 6.2   & $\pm$ & 44.9  &  456.6 & $\pm$ & 5.1   & $\pm$ & 31.4  \\
 0.750 &  291.5 & $\pm$ & 9.5  & $\pm$ & 16.8 &  340.6 & $\pm$ &  6.9  & $\pm$ & 20.3 &  256.3 & $\pm$ & 4.6   & $\pm$ & 21.6  &  228.6 & $\pm$ & 4.0   & $\pm$ & 15.8  \\
 0.864 &  198.9 & $\pm$ & 5.5  & $\pm$ & 11.7 &  215.4 & $\pm$ &  3.8  & $\pm$ & 12.8 &  143.3 & $\pm$ & 2.4   & $\pm$ & 12.1  &  105.6 & $\pm$ & 1.8   & $\pm$ & 7.3   \\
 0.950 &  155.2 & $\pm$ & 4.1  & $\pm$ & 9.3  &  179.1 & $\pm$ &  2.8  & $\pm$ & 10.7 &  109.2 & $\pm$ & 1.7   & $\pm$ & 9.2   &  80.01 & $\pm$ & 1.28  & $\pm$ & 5.52  \\
 1.052 &  104.4 & $\pm$ & 3.9  & $\pm$ & 6.4  &  121.1 & $\pm$ &  2.6  & $\pm$ & 7.2  &  80.57 & $\pm$ & 1.69  & $\pm$ & 6.78  &  54.58 & $\pm$ & 1.20  & $\pm$ & 3.77  \\
 1.142 &  72.04 & $\pm$ & 3.15 & $\pm$ & 4.47 &  84.11 & $\pm$ &  2.12 & $\pm$ & 5.02 &  54.95 & $\pm$ & 1.35  & $\pm$ & 4.62  &  35.43 & $\pm$ & 0.92  & $\pm$ & 2.45  \\
 1.254 &  52.80 & $\pm$ & 2.91 & $\pm$ & 3.36 &  61.08 & $\pm$ &  1.88 & $\pm$ & 3.65 &  39.10 & $\pm$ & 1.20  & $\pm$ & 3.29  &  22.99 & $\pm$ & 0.77  & $\pm$ & 1.59  \\
 1.355 &  43.36 & $\pm$ & 2.73 & $\pm$ & 2.83 &  51.28 & $\pm$ &  1.75 & $\pm$ & 3.07 &  28.43 & $\pm$ & 1.06  & $\pm$ & 2.39  &  16.49 & $\pm$ & 0.66  & $\pm$ & 1.14  \\
 1.449 &  33.71 & $\pm$ & 2.46 & $\pm$ & 2.25 &  39.52 & $\pm$ &  1.54 & $\pm$ & 2.36 &  22.06 & $\pm$ & 0.86  & $\pm$ & 1.85  &  11.80 & $\pm$ & 0.51  & $\pm$ & 0.82  \\
 1.548 &  29.01 & $\pm$ & 2.52 & $\pm$ & 1.99 &  31.39 & $\pm$ &  1.47 & $\pm$ & 1.88 &  16.45 & $\pm$ & 0.82  & $\pm$ & 1.38  &  8.36  & $\pm$ & 0.46  & $\pm$ & 0.58  \\
 1.648 &  20.53 & $\pm$ & 2.18 & $\pm$ & 1.44 &  22.11 & $\pm$ &  1.26 & $\pm$ & 1.32 &  12.12 & $\pm$ & 0.73  & $\pm$ & 1.02  &  6.09  & $\pm$ & 0.40  & $\pm$ & 0.42  \\
 1.751 &  16.39 & $\pm$ & 2.15 & $\pm$ & 1.18 &  18.51 & $\pm$ &  1.21 & $\pm$ & 1.11 &  10.03 & $\pm$ & 0.70  & $\pm$ & 0.84  &  4.31  & $\pm$ & 0.35  & $\pm$ & 0.30  \\
 1.845 &  14.62 & $\pm$ & 2.56 & $\pm$ & 1.08 &  14.79 & $\pm$ &  1.16 & $\pm$ & 0.89 &  8.13  & $\pm$ & 0.67  & $\pm$ & 0.68  &  2.64  & $\pm$ & 0.29  & $\pm$ & 0.18  \\
 1.959 &  10.02 & $\pm$ & 2.08 & $\pm$ & 0.76 &  9.43  & $\pm$ &  0.99 & $\pm$ & 0.57 &  6.13  & $\pm$ & 0.64  & $\pm$ & 0.51  &  1.69  & $\pm$ & 0.27  & $\pm$ & 0.12  \\
 2.051 &  6.74  & $\pm$ & 1.83 & $\pm$ & 0.53 &  8.92  & $\pm$ &  0.96 & $\pm$ & 0.54 &  3.97  & $\pm$ & 0.52  & $\pm$ & 0.33  &  1.58  & $\pm$ & 0.25  & $\pm$ & 0.11  \\
 2.147 &  15.29 & $\pm$ & 4.62 & $\pm$ & 1.23 &  6.80  & $\pm$ &  0.87 & $\pm$ & 0.41 &  3.93  & $\pm$ & 0.56  & $\pm$ & 0.33  &  1.70  & $\pm$ & 0.27  & $\pm$ & 0.12  \\
 2.250 &  10.37 & $\pm$ & 4.24 & $\pm$ & 0.86 &  4.75  & $\pm$ &  0.75 & $\pm$ & 0.29 &  2.37  & $\pm$ & 0.47  & $\pm$ & 0.20  &  0.779 & $\pm$ & 0.187 & $\pm$ & 0.055 \\
 2.331 &  5.44  & $\pm$ & 2.72 & $\pm$ & 0.46 &  4.45  & $\pm$ &  0.87 & $\pm$ & 0.27 &  2.88  & $\pm$ & 0.76  & $\pm$ & 0.24  &  1.39  & $\pm$ & 0.34  & $\pm$ & 0.10  \\
 2.458 &  6.94  & $\pm$ & 4.95 & $\pm$ & 0.60 &  3.98  & $\pm$ &  1.33 & $\pm$ & 0.24 &  1.27  & $\pm$ & 0.63  & $\pm$ & 0.11  &  0.46  & $\pm$ & 0.265 & $\pm$ & 0.032 \\
 2.548 &  3.14  & $\pm$ & 3.15 & $\pm$ & 0.28 &  3.96  & $\pm$ &  1.32 & $\pm$ & 0.24 &  0.989 & $\pm$ & 0.572 & $\pm$ & 0.083 &        &       &       &       &       \\
 2.651 &  6.27  & $\pm$ & 4.45 & $\pm$ & 0.57 &  3.43  & $\pm$ &  1.35 & $\pm$ & 0.21 &  2.15  & $\pm$ & 0.81  & $\pm$ & 0.18  &  0.309 & $\pm$ & 0.219 & $\pm$ & 0.022 \\
 2.749 &        &       &      &       &      &  3.28  & $\pm$ &  1.50 & $\pm$ & 0.20 &  1.07  & $\pm$ & 0.62  & $\pm$ & 0.09  &  0.349 & $\pm$ & 0.248 & $\pm$ & 0.025 \\
 2.886 &  5.86  & $\pm$ & 5.88 & $\pm$ & 0.57 &  7.14  & $\pm$ &  2.34 & $\pm$ & 0.43 &  0.748 & $\pm$ & 0.434 & $\pm$ & 0.063 &  0.503 & $\pm$ & 0.254 & $\pm$ & 0.036 \\ 
\hline 
\hline\hline
 &\multicolumn{5}{c|}{}  
 &\multicolumn{5}{c|}{}  
 &\multicolumn{5}{c|}{}  
 &\multicolumn{5}{c|}{}
 \\ 
 $\langle E_\gamma \rangle $ (GeV) 
 &\multicolumn{5}{c|}{$50^\circ \leq \theta_p^{\rm{c.m.}} <60^\circ$} 
 &\multicolumn{5}{c|}{$60^\circ \leq \theta_p^{\rm{c.m.}} <70^\circ$} 
 &\multicolumn{5}{c|}{$70^\circ \leq \theta_p^{\rm{c.m.}} <80^\circ$}
 &\multicolumn{5}{c|}{$80^\circ \leq \theta_p^{\rm{c.m.}} <90$}   \\ 
 &\multicolumn{5}{c|}{}  
 &\multicolumn{5}{c|}{}  
 &\multicolumn{5}{c|}{}  
 &\multicolumn{5}{c|}{}
\\
\hline
 0.560 &  632.3 & $\pm$ & 6.0  & $\pm$ & 33.0  &  637.9 & $\pm$ & 5.4   & $\pm$ & 26.8  &  608.7 & $\pm$ & 5.3   & $\pm$ & 26.4  &  608.8 & $\pm$ & 5.1   & $\pm$ & 19.3  \\
 0.641 &  426.2 & $\pm$ & 4.0  & $\pm$ & 22.3  &  441.8 & $\pm$ & 3.7   & $\pm$ & 18.6  &  426.6 & $\pm$ & 3.5   & $\pm$ & 18.5  &  413.9 & $\pm$ & 3.4   & $\pm$ & 13.2  \\
 0.750 &  207.3 & $\pm$ & 3.1  & $\pm$ & 10.9  &  215.5 & $\pm$ & 2.7   & $\pm$ & 9.1   &  208.9 & $\pm$ & 2.6   & $\pm$ & 9.1   &  211.3 & $\pm$ & 2.6   & $\pm$ & 6.8   \\
 0.864 &  87.07 & $\pm$ & 1.38 & $\pm$ & 4.61  &  87.84 & $\pm$ & 1.16  & $\pm$ & 3.72  &  89.08 & $\pm$ & 1.23  & $\pm$ & 3.90  &  86.93 & $\pm$ & 1.10  & $\pm$ & 2.81  \\
 0.950 &  56.05 & $\pm$ & 0.90 & $\pm$ & 2.98  &  50.54 & $\pm$ & 0.71  & $\pm$ & 2.15  &  52.91 & $\pm$ & 0.76  & $\pm$ & 2.32  &  52.75 & $\pm$ & 0.69  & $\pm$ & 1.72  \\
 1.052 &  38.11 & $\pm$ & 0.85 & $\pm$ & 2.04  &  29.38 & $\pm$ & 0.62  & $\pm$ & 1.25  &  31.05 & $\pm$ & 0.66  & $\pm$ & 1.37  &  30.14 & $\pm$ & 0.59  & $\pm$ & 0.99  \\
 1.142 &  25.56 & $\pm$ & 0.66 & $\pm$ & 1.38  &  18.68 & $\pm$ & 0.48  & $\pm$ & 0.8   &  19.53 & $\pm$ & 0.50  & $\pm$ & 0.86  &  17.41 & $\pm$ & 0.43  & $\pm$ & 0.58  \\
 1.254 &  15.05 & $\pm$ & 0.53 & $\pm$ & 0.82  &  10.29 & $\pm$ & 0.37  & $\pm$ & 0.44  &  10.10 & $\pm$ & 0.37  & $\pm$ & 0.45  &  9.63  & $\pm$ & 0.33  & $\pm$ & 0.32  \\
 1.355 &  9.77  & $\pm$ & 0.43 & $\pm$ & 0.54  &  7.35  & $\pm$ & 0.33  & $\pm$ & 0.32  &  6.29  & $\pm$ & 0.30  & $\pm$ & 0.28  &  7.19  & $\pm$ & 0.29  & $\pm$ & 0.24  \\
 1.449 &  7.12  & $\pm$ & 0.37 & $\pm$ & 0.39  &  4.68  & $\pm$ & 0.26  & $\pm$ & 0.2   &  4.62  & $\pm$ & 0.23  & $\pm$ & 0.21  &  4.63  & $\pm$ & 0.23  & $\pm$ & 0.16  \\
 1.548 &  5.62  & $\pm$ & 0.35 & $\pm$ & 0.31  &  4.35  & $\pm$ & 0.28  & $\pm$ & 0.19  &  3.95  & $\pm$ & 0.23  & $\pm$ & 0.18  &  3.42  & $\pm$ & 0.21  & $\pm$ & 0.12  \\
 1.648 &  3.80  & $\pm$ & 0.29 & $\pm$ & 0.21  &  3.49  & $\pm$ & 0.26  & $\pm$ & 0.15  &  2.90  & $\pm$ & 0.20  & $\pm$ & 0.13  &  2.83  & $\pm$ & 0.19  & $\pm$ & 0.10  \\
 1.751 &  3.07  & $\pm$ & 0.27 & $\pm$ & 0.18  &  3.04  & $\pm$ & 0.25  & $\pm$ & 0.13  &  1.98  & $\pm$ & 0.17  & $\pm$ & 0.09  &  2.10  & $\pm$ & 0.17  & $\pm$ & 0.07  \\
 1.845 &  2.20  & $\pm$ & 0.25 & $\pm$ & 0.13  &  1.92  & $\pm$ & 0.22  & $\pm$ & 0.09  &  1.58  & $\pm$ & 0.16  & $\pm$ & 0.07  &  1.32  & $\pm$ & 0.14  & $\pm$ & 0.05  \\
 1.959 &  2.17  & $\pm$ & 0.28 & $\pm$ & 0.13  &  1.01  & $\pm$ & 0.19  & $\pm$ & 0.05  &  1.15  & $\pm$ & 0.15  & $\pm$ & 0.05  &  1.19  & $\pm$ & 0.15  & $\pm$ & 0.04  \\
 2.051 &  1.08  & $\pm$ & 0.18 & $\pm$ & 0.06  &  1.05  & $\pm$ & 0.18  & $\pm$ & 0.05  &  0.735 & $\pm$ & 0.117 & $\pm$ & 0.034 &  0.551 & $\pm$ & 0.099 & $\pm$ & 0.02  \\
 2.147 &  1.07  & $\pm$ & 0.19 & $\pm$ & 0.06  &  0.95  & $\pm$ & 0.168 & $\pm$ & 0.043 &  0.830 & $\pm$ & 0.211 & $\pm$ & 0.038 &  0.533 & $\pm$ & 0.099 & $\pm$ & 0.02  \\
 2.250 &  0.507 & $\pm$ & 0.144& $\pm$ & 0.031 &  0.661 & $\pm$ & 0.146 & $\pm$ & 0.03  &  0.361 & $\pm$ & 0.116 & $\pm$ & 0.017 &  0.503 & $\pm$ & 0.118 & $\pm$ & 0.019 \\
 2.331 &  0.970 & $\pm$ & 0.281& $\pm$ & 0.059 &  0.647 & $\pm$ & 0.180 & $\pm$ & 0.029 &  0.408 & $\pm$ & 0.152 & $\pm$ & 0.019 &  0.36  & $\pm$ & 0.149 & $\pm$ & 0.014 \\
 2.458 &  0.795 & $\pm$ & 0.326& $\pm$ & 0.049 &  0.125 & $\pm$ & 0.125 & $\pm$ & 0.006 &  0.195 & $\pm$ & 0.138 & $\pm$ & 0.009 &  0.131 & $\pm$ & 0.093 & $\pm$ & 0.005 \\
 2.548 &  0.785 & $\pm$ & 0.321& $\pm$ & 0.049 &  0.121 & $\pm$ & 0.121 & $\pm$ & 0.006 &  0.072 & $\pm$ & 0.072 & $\pm$ & 0.003 &        &       &       &       &       \\
 2.651 &  0.257 & $\pm$ & 0.182& $\pm$ & 0.016 &  0.315 & $\pm$ & 0.183 & $\pm$ & 0.015 &  0.183 & $\pm$ & 0.134 & $\pm$ & 0.009 &  0.063 & $\pm$ & 0.063 & $\pm$ & 0.002 \\
 2.749 &  0.428 & $\pm$ & 0.248& $\pm$ & 0.028 &  0.237 & $\pm$ & 0.168 & $\pm$ & 0.011 &        &       &       &       &       &  0.148 & $\pm$ & 0.105 & $\pm$ & 0.006 \\
 2.886 &  0.224 & $\pm$ & 0.225& $\pm$ & 0.015 &  0.134 & $\pm$ & 0.134 & $\pm$ & 0.006 &  0.115 & $\pm$ & 0.115 & $\pm$ & 0.005 &  0.110 & $\pm$ & 0.110 & $\pm$ & 0.004 \\
 \hline
 \end{tabular}
 \end{center}
 \caption{ \small Differential cross sections $d\sigma/d\Omega$ in nb/sterad of the deuteron photodisintegration for photon energies 0.5-3.0~GeV
and for center-of-mass proton angles $10^\circ \leq \vartheta_p^{\rm{c.m.}} < 90^\circ$. The first error is statistical and the second is systematic 
in each case.}
 \label{tab:data2}
 \end{table*}
  
%%%%%	aaaaaa  
 \begin{table*}[htbp]
 \begin{center}
 \begin{tabular}{|c|ccccc|ccccc|ccccc|ccccc|} \hline\hline  
  &\multicolumn{5}{c|}{}  
  &\multicolumn{5}{c|}{}  
  &\multicolumn{5}{c|}{}  
  &\multicolumn{5}{c|}{}
 \\           
 $\langle E_\gamma \rangle $ (GeV)
 &\multicolumn{5}{c|}{$90^\circ \leq \theta_p^{\rm{c.m.}} <100^\circ$}
 &\multicolumn{5}{c|}{$100^\circ \leq \theta_p^{\rm{c.m.}} <110^\circ$} 
 &\multicolumn{5}{c|}{$110^\circ \leq \theta_p^{\rm{c.m.}} <120^\circ$} 
 &\multicolumn{5}{c|}{$120^\circ \leq \theta_p^{\rm{c.m.}} <130$}
 \\ 
 &\multicolumn{5}{c|}{}  
 &\multicolumn{5}{c|}{}  
 &\multicolumn{5}{c|}{}  
 &\multicolumn{5}{c|}{}
\\
\hline
 0.560 &  595.8 & $\pm$ & 5.1   & $\pm$ & 23.8 &  545.4 & $\pm$ & 5.0   & $\pm$ & 22.3  &  477.9 & $\pm$ & 6.9   & $\pm$ & 25.9  &  407.8  & $\pm$ & 7.1   &  $\pm$ & 23.9 \\
 0.641 &  394.9 & $\pm$ & 3.6   & $\pm$ & 15.9 &  358.6 & $\pm$ & 3.1   & $\pm$ & 14.7  &  322.1 & $\pm$ & 3.4   & $\pm$ & 17.5  &  258.0  & $\pm$ & 3.5   &  $\pm$ & 15.2 \\
 0.750 &  209.3 & $\pm$ & 2.8   & $\pm$ & 8.6  &  185.3 & $\pm$ & 2.4   & $\pm$ & 7.7   &  171.2 & $\pm$ & 2.6   & $\pm$ & 9.4   &  145.0  & $\pm$ & 2.8   &  $\pm$ & 8.6  \\
 0.864 &  86.76 & $\pm$ & 1.18  & $\pm$ & 3.62 &  85.01 & $\pm$ & 1.10  & $\pm$ & 3.57  &  77.02 & $\pm$ & 1.16  & $\pm$ & 4.26  &  57.99  & $\pm$ & 1.18  &  $\pm$ & 3.49 \\
 0.950 &  50.64 & $\pm$ & 0.72  & $\pm$ & 2.16 &  53.61 & $\pm$ & 0.70  & $\pm$ & 2.27  &  47.33 & $\pm$ & 0.73  & $\pm$ & 2.64  &  37.02  & $\pm$ & 0.68  &  $\pm$ & 2.25 \\
 1.052 &  31.12 & $\pm$ & 0.59  & $\pm$ & 1.36 &  30.12 & $\pm$ & 0.59  & $\pm$ & 1.29  &  29.15 & $\pm$ & 0.59  & $\pm$ & 1.64  &  21.77  & $\pm$ & 0.58  &  $\pm$ & 1.35 \\
 1.142 &  17.30 & $\pm$ & 0.41  & $\pm$ & 0.77 &  18.01 & $\pm$ & 0.43  & $\pm$ & 0.78  &  17.09 & $\pm$ & 0.42  & $\pm$ & 0.97  &  13.27  & $\pm$ & 0.43  &  $\pm$ & 0.83 \\
 1.254 &  10.43 & $\pm$ & 0.33  & $\pm$ & 0.48 &  10.55 & $\pm$ & 0.34  & $\pm$ & 0.47  &  8.62  & $\pm$ & 0.31  & $\pm$ & 0.50  &  7.49   & $\pm$ & 0.33  &  $\pm$ & 0.48 \\
 1.355 &  6.36  & $\pm$ & 0.26  & $\pm$ & 0.30 &  6.48  & $\pm$ & 0.27  & $\pm$ & 0.29  &  5.23  & $\pm$ & 0.24  & $\pm$ & 0.31  &  5.21   & $\pm$ & 0.27  &  $\pm$ & 0.34 \\
 1.449 &  3.69  & $\pm$ & 0.20  & $\pm$ & 0.18 &  3.79  & $\pm$ & 0.20  & $\pm$ & 0.17  &  3.45  & $\pm$ & 0.19  & $\pm$ & 0.21  &  3.17   & $\pm$ & 0.21  &  $\pm$ & 0.21 \\
 1.548 &  3.13  & $\pm$ & 0.19  & $\pm$ & 0.16 &  3.02  & $\pm$ & 0.19  & $\pm$ & 0.14  &  2.17  & $\pm$ & 0.16  & $\pm$ & 0.13  &  2.53   & $\pm$ & 0.20  &  $\pm$ & 0.17 \\
 1.648 &  2.17  & $\pm$ & 0.16  & $\pm$ & 0.11 &  1.83  & $\pm$ & 0.15  & $\pm$ & 0.09  &  1.94  & $\pm$ & 0.15  & $\pm$ & 0.12  &  1.97   & $\pm$ & 0.17  &  $\pm$ & 0.14 \\
 1.751 &  1.36  & $\pm$ & 0.13  & $\pm$ & 0.07 &  1.39  & $\pm$ & 0.13  & $\pm$ & 0.07  &  1.27  & $\pm$ & 0.13  & $\pm$ & 0.08  &  1.18   & $\pm$ & 0.13  &  $\pm$ & 0.09 \\
 1.845 &  1.03  & $\pm$ & 0.12  & $\pm$ & 0.06 &  0.694 & $\pm$ & 0.099 & $\pm$ & 0.034 &  1.22  & $\pm$ & 0.13  & $\pm$ & 0.08  &  1.11   & $\pm$ & 0.14  &  $\pm$ & 0.08 \\
 1.959 &  0.633 & $\pm$ & 0.109 & $\pm$ & 0.037&  0.748 & $\pm$ & 0.109 & $\pm$ & 0.037 &  0.599 & $\pm$ & 0.103 & $\pm$ & 0.039 &  0.719  & $\pm$ & 0.111 &  $\pm$ & 0.055\\
 2.051 &  0.600 & $\pm$ & 0.105 & $\pm$ & 0.036&  0.643 & $\pm$ & 0.100 & $\pm$ & 0.032 &  0.638 & $\pm$ & 0.111 & $\pm$ & 0.042 &  0.456  & $\pm$ & 0.086 &  $\pm$ & 0.036\\
 2.147 &  0.310 & $\pm$ & 0.109 & $\pm$ & 0.019&  0.310 & $\pm$ & 0.072 & $\pm$ & 0.016 &  0.479 & $\pm$ & 0.134 & $\pm$ & 0.032 &  0.562  & $\pm$ & 0.102 &  $\pm$ & 0.045\\
 2.250 &  0.766 & $\pm$ & 0.203 & $\pm$ & 0.049&  0.304 & $\pm$ & 0.076 & $\pm$ & 0.016 &  0.420 & $\pm$ & 0.086 & $\pm$ & 0.029 &  0.284  & $\pm$ & 0.072 &  $\pm$ & 0.024\\
 2.331 &  0.297 & $\pm$ & 0.094 & $\pm$ & 0.019&  0.367 & $\pm$ & 0.123 & $\pm$ & 0.019 &  0.243 & $\pm$ & 0.115 & $\pm$ & 0.017 &  0.403  & $\pm$ & 0.107 &  $\pm$ & 0.034\\
 2.458 &  0.250 & $\pm$ & 0.125 & $\pm$ & 0.017&  0.116 & $\pm$ & 0.083 & $\pm$ & 0.006 &  0.236 & $\pm$ & 0.118 & $\pm$ & 0.017 &  0.117  & $\pm$ & 0.083 &  $\pm$ & 0.010\\
 2.548 &  0.134 & $\pm$ & 0.095 & $\pm$ & 0.009&  0.188 & $\pm$ & 0.110 & $\pm$ & 0.010 &  0.058 & $\pm$ & 0.058 & $\pm$ & 0.004 &  0.191  & $\pm$ & 0.112 &  $\pm$ & 0.017\\
 2.651 &  0.181 & $\pm$ & 0.105 & $\pm$ & 0.013&  0.056 & $\pm$ & 0.056 & $\pm$ & 0.003 &  0.162 & $\pm$ & 0.094 & $\pm$ & 0.012 &  0.214  & $\pm$ & 0.108 &  $\pm$ & 0.020\\
 2.749 &  0.139 & $\pm$ & 0.098 & $\pm$ & 0.010&        &       &       &       &       &  0.199 & $\pm$ & 0.115 & $\pm$ & 0.015 &  0.123  & $\pm$ & 0.087 &  $\pm$ & 0.012\\
 2.886 &  0.230 & $\pm$ & 0.164 & $\pm$ & 0.018&  0.066 & $\pm$ & 0.066 & $\pm$ & 0.004 &  0.066 & $\pm$ & 0.066 & $\pm$ & 0.005 &  0.063  & $\pm$ & 0.063 &  $\pm$ & 0.006\\
 \hline 
 \hline 
 \hline
   &\multicolumn{5}{c|}{}  
   &\multicolumn{5}{c|}{}  
   &\multicolumn{5}{c|}{}  
   &\multicolumn{5}{c|}{}  
   \\   
  $\langle E_\gamma \rangle $ (GeV)
   & \multicolumn{5}{c|}{$130^\circ \leq \theta_p^{\rm{c.m.}} <140^\circ$} 
   & \multicolumn{5}{c|}{$140^\circ \leq \theta_p^{\rm{c.m.}} <150^\circ$}
   & \multicolumn{5}{c|}{$150^\circ \leq \theta_p^{\rm{c.m.}} <160$}
   & \multicolumn{5}{c|}{}
 \\
   &\multicolumn{5}{c|}{}  
   &\multicolumn{5}{c|}{}  
   &\multicolumn{5}{c|}{}  
   &\multicolumn{5}{c|}{} 
  \\
 \hline
 0.560 &  383.2  & $\pm$  &5.5   & $\pm$  & 18.0  &  347.4   &$\pm$  & 12.1   & $\pm$ & 15.4  &  149.2  &  $\pm$  &  15.6   &  $\pm$   & 10.8  &&&&&\\
 0.641 &  231.3  & $\pm$  &3.2   & $\pm$  & 10.9  &  207.7   &$\pm$  & 4.0    & $\pm$ & 9.2   &  73.75  &  $\pm$  &  5.31   &  $\pm$   & 5.33  &&&&&\\
 0.750 &  126.4  & $\pm$  &2.5   & $\pm$  & 6.0   &  120.7   &$\pm$  & 3.2    & $\pm$ & 5.4   &  53.32  &  $\pm$  &  4.45   &  $\pm$   & 3.85  &&&&&\\
 0.864 &  59.20  & $\pm$  &1.14  & $\pm$  & 2.80  &  56.80   &$\pm$  & 1.37   & $\pm$ & 2.58  &  41.52  &  $\pm$  &  2.42   &  $\pm$   & 3.00  &&&&&\\
 0.950 &  38.31  & $\pm$  &0.74  & $\pm$  & 1.81  &  39.69   &$\pm$  & 0.89   & $\pm$ & 1.81  &  29.99  &  $\pm$  &  1.44   &  $\pm$   & 2.17  &&&&&\\
 1.052 &  23.56  & $\pm$  &0.65  & $\pm$  & 1.12  &  27.65   &$\pm$  & 0.82   & $\pm$ & 1.27  &  19.28  &  $\pm$  &  1.13   &  $\pm$   & 1.39  &&&&&\\
 1.142 &  13.28  & $\pm$  &0.46  & $\pm$  & 0.63  &  18.36   &$\pm$  & 0.61   & $\pm$ & 0.85  &  17.35  &  $\pm$  &  0.96   &  $\pm$   & 1.25  &&&&&\\
 1.254 &  7.44   & $\pm$  &0.35  & $\pm$  & 0.35  &  14.58   &$\pm$  & 0.55   & $\pm$ & 0.68  &  13.96  &  $\pm$  &  0.84   &  $\pm$   & 1.01  &&&&&\\
 1.355 &  5.90   & $\pm$  &0.31  & $\pm$  & 0.28  &  12.09   &$\pm$  & 0.49   & $\pm$ & 0.57  &  13.15  &  $\pm$  &  0.80   &  $\pm$   & 0.95  &&&&&\\
 1.449 &  4.08   & $\pm$  &0.25  & $\pm$  & 0.20  &  9.34    &$\pm$  & 0.42   & $\pm$ & 0.45  &  13.71  &  $\pm$  &  0.79   &  $\pm$   & 0.99  &&&&&\\
 1.548 &  2.81   & $\pm$  &0.22  & $\pm$  & 0.13  &  7.29    &$\pm$  & 0.39   & $\pm$ & 0.35  &  10.50  &  $\pm$  &  0.70   &  $\pm$   & 0.76  &&&&&\\
 1.648 &  1.59   & $\pm$  &0.15  & $\pm$  & 0.08  &  4.66    &$\pm$  & 0.30   & $\pm$ & 0.23  &  8.05   &  $\pm$  &  0.59   &  $\pm$   & 0.58  &&&&&\\
 1.751 &  1.20   & $\pm$  &0.13  & $\pm$  & 0.06  &  3.33    &$\pm$  & 0.26   & $\pm$ & 0.16  &  5.99   &  $\pm$  &  0.51   &  $\pm$   & 0.43  &&&&&\\
 1.845 &  1.07   & $\pm$  &0.14  & $\pm$  & 0.05  &  2.55    &$\pm$  & 0.24   & $\pm$ & 0.13  &  4.90   &  $\pm$  &  0.48   &  $\pm$   & 0.35  &&&&&\\
 1.959 &  0.692  & $\pm$  &0.114 & $\pm$  & 0.034 &  1.15    &$\pm$  & 0.16   & $\pm$ & 0.06  &  2.49   &  $\pm$  &  0.35   &  $\pm$   & 0.18  &&&&&\\
 2.051 &  0.507  & $\pm$  &0.110 & $\pm$  & 0.025 &  1.33    &$\pm$  & 0.17   & $\pm$ & 0.07  &  3.10   &  $\pm$  &  0.38   &  $\pm$   & 0.22  &&&&&\\
 2.147 &  0.320  & $\pm$  &0.080 & $\pm$  & 0.016 &  0.942   &$\pm$  & 0.146  & $\pm$ & 0.048 &  2.38   &  $\pm$  &  0.34   &  $\pm$   & 0.17  &&&&&\\
 2.250 &  0.555  & $\pm$  &0.134 & $\pm$  & 0.027 &  0.552   &$\pm$  & 0.113  & $\pm$ & 0.028 &  1.62   &  $\pm$  &  0.28   &  $\pm$   & 0.12  &&&&&\\
 2.331 &  0.481  & $\pm$  &0.139 & $\pm$  & 0.023 &  0.531   &$\pm$  & 0.161  & $\pm$ & 0.027 &  1.05   &  $\pm$  &  0.29   &  $\pm$   & 0.08  &&&&&\\
 2.458 &  0.166  & $\pm$  &0.096 & $\pm$  & 0.008 &  0.364   &$\pm$  & 0.164  & $\pm$ & 0.019 &  0.832  &  $\pm$  &  0.343  &  $\pm$   & 0.06  &&&&&\\
 2.548 &  0.327  & $ \pm$ &0.135 & $\pm$  & 0.016 &  0.278   &$\pm$  & 0.140  & $\pm$ & 0.015 &  1.08   &  $\pm$  &  0.39   &  $\pm$   & 0.08  &&&&&\\
 2.651 &         &        &      &        &       &  0.310   &$\pm$  & 0.140  & $\pm$ & 0.017 &  0.325  &  $\pm$  &  0.189  &  $\pm$   & 0.023 &&&&&\\
 2.749 &  0.166  &  $\pm$ & 0.096& $\pm$  & 0.008 &  0.136   &$\pm$  & 0.097  & $\pm$ & 0.007 &  0.535  &  $\pm$  &  0.276  &  $\pm$   & 0.039 &&&&&\\
 2.886 &  0.064  &  $\pm$ & 0.047& $\pm$  & 0.003 &  0.072   &$\pm$  & 0.073  & $\pm$ & 0.004 &  0.516  &  $\pm$  &  0.197  &  $\pm$   & 0.037 &&&&&\\ \hline
 \end{tabular}
 \end{center}
 \caption{ \small Differential cross sections $d\sigma/d\Omega$ in nb/sterad of the deuteron photodisintegration for photon energies 0.5-3.0~GeV
and for center-of-mass proton angles $90^\circ \leq \vartheta_p^{\rm{c.m.}} < 160^\circ$. The first error is statistical and the second is systematic 
in each case.}
 \label{tab:data4}
 \end{table*}
\bibliography{deuteron_PRC}

\begin{thebibliography}{29}
\expandafter\ifx\csname natexlab\endcsname\relax\def\natexlab#1{#1}\fi
\expandafter\ifx\csname bibnamefont\endcsname\relax
  \def\bibnamefont#1{#1}\fi
\expandafter\ifx\csname bibfnamefont\endcsname\relax
  \def\bibfnamefont#1{#1}\fi
\expandafter\ifx\csname citenamefont\endcsname\relax
  \def\citenamefont#1{#1}\fi
\expandafter\ifx\csname url\endcsname\relax
  \def\url#1{\texttt{#1}}\fi
\expandafter\ifx\csname urlprefix\endcsname\relax\def\urlprefix{URL }\fi
\providecommand{\bibinfo}[2]{#2}
\providecommand{\eprint}[2][]{\url{#2}}

\bibitem[{\citenamefont{Gilman and Gross}(2002)}]{GilGro}
\bibinfo{author}{\bibfnamefont{R.}~\bibnamefont{Gilman}} \bibnamefont{and}
  \bibinfo{author}{\bibfnamefont{F.}~\bibnamefont{Gross}}, \bibinfo{journal}{J.
  Phys. G} \textbf{\bibinfo{volume}{R37}}, \bibinfo{pages}{28}
  (\bibinfo{year}{2002}).

\bibitem[{\citenamefont{Holt}(1990)}]{holt}
\bibinfo{author}{\bibfnamefont{R.~J.} \bibnamefont{Holt}},
  \bibinfo{journal}{Phys.\ Rev.} \textbf{\bibinfo{volume}{C41}},
  \bibinfo{pages}{2400} (\bibinfo{year}{1990}).

\bibitem[{\citenamefont{{Matveev, R. M. Muradyan and A. N.
  Tavkhelidze}}(1973)}]{Matveev}
\bibinfo{author}{\bibfnamefont{V.~A.} \bibnamefont{{Matveev, R. M. Muradyan and
  A. N. Tavkhelidze}}}, \bibinfo{journal}{Nuovo\ Cimento}
  \textbf{\bibinfo{volume}{7}}, \bibinfo{pages}{719} (\bibinfo{year}{1973}).

\bibitem[{\citenamefont{Brodsky and Farrar}(1973)}]{Brodsky}
\bibinfo{author}{\bibfnamefont{S.~J.} \bibnamefont{Brodsky}} \bibnamefont{and}
  \bibinfo{author}{\bibfnamefont{G.}~\bibnamefont{Farrar}},
  \bibinfo{journal}{Phys.\ Rev.\ Lett.} \textbf{\bibinfo{volume}{31}},
  \bibinfo{pages}{1153} (\bibinfo{year}{1973}).

\bibitem[{\citenamefont{{Napolitano \emph{et al.}}}(1988)}]{SLACNE81}
\bibinfo{author}{\bibfnamefont{J.}~\bibnamefont{{Napolitano \emph{et al.}}}},
  \bibinfo{journal}{Phys.\ Rev.\ Lett.} \textbf{\bibinfo{volume}{61}},
  \bibinfo{pages}{2530} (\bibinfo{year}{1988}).

\bibitem[{\citenamefont{{Freedman \emph{et al.}}}(1993)}]{SLACNE82}
\bibinfo{author}{\bibfnamefont{S.~J.} \bibnamefont{{Freedman \emph{et al.}}}},
  \bibinfo{journal}{Phys.\ Rev.} \textbf{\bibinfo{volume}{C48}},
  \bibinfo{pages}{1864} (\bibinfo{year}{1993}).

\bibitem[{\citenamefont{{Belz \emph{et al.}}}(1995)}]{SLACNE17}
\bibinfo{author}{\bibfnamefont{J.~E.} \bibnamefont{{Belz \emph{et al.}}}},
  \bibinfo{journal}{Phys.\ Rev.\ Lett.} \textbf{\bibinfo{volume}{74}},
  \bibinfo{pages}{646} (\bibinfo{year}{1995}).

\bibitem[{\citenamefont{{Bochna \emph{et al.}}}(1998)}]{Bochna}
\bibinfo{author}{\bibfnamefont{C.}~\bibnamefont{{Bochna \emph{et al.}}}},
  \bibinfo{journal}{Phys.\ Rev.\ Lett.} \textbf{\bibinfo{volume}{81}},
  \bibinfo{pages}{4576} (\bibinfo{year}{1998}).

\bibitem[{\citenamefont{{Schulte \emph{et al.}}}(2001)}]{Schulte}
\bibinfo{author}{\bibfnamefont{E.~C.} \bibnamefont{{Schulte \emph{et al.}}}},
  \bibinfo{journal}{Phys.\ Rev.\ Lett.} \textbf{\bibinfo{volume}{87}},
  \bibinfo{pages}{102302} (\bibinfo{year}{2001}).

\bibitem[{\citenamefont{{Schulte \emph{et al.}}}(2002)}]{SchulteA}
\bibinfo{author}{\bibfnamefont{E.~C.} \bibnamefont{{Schulte \emph{et al.}}}},
  \bibinfo{journal}{Phys.\ Rev.} \textbf{\bibinfo{volume}{C66}},
  \bibinfo{pages}{0422011} (\bibinfo{year}{2002}).

\bibitem[{\citenamefont{{Wijesooriya \emph{et al.}}}(2001)}]{pol}
\bibinfo{author}{\bibfnamefont{K.}~\bibnamefont{{Wijesooriya \emph{et al.}}}},
  \bibinfo{journal}{Phys.\ Rev.\ Lett.} \textbf{\bibinfo{volume}{86}},
  \bibinfo{pages}{2975} (\bibinfo{year}{2001}).

\bibitem[{\citenamefont{{Adamian \emph{et al.}}}(1998)}]{adam}
\bibinfo{author}{\bibfnamefont{F.~V.} \bibnamefont{{Adamian \emph{et al.}}}},
  \bibinfo{journal}{Nucl.\ Phys.} \textbf{\bibinfo{volume}{14}},
  \bibinfo{pages}{831} (\bibinfo{year}{1998}).

\bibitem[{\citenamefont{Brodsky and Hiller}(1983)}]{rna}
\bibinfo{author}{\bibfnamefont{S.~L.} \bibnamefont{Brodsky}} \bibnamefont{and}
  \bibinfo{author}{\bibfnamefont{J.~R.} \bibnamefont{Hiller}},
  \bibinfo{journal}{Phys.\ Rev.} \textbf{\bibinfo{volume}{C28}},
  \bibinfo{pages}{475} (\bibinfo{year}{1983}).

\bibitem[{\citenamefont{{Frankfurt, G.A. MIller, M.M. Sargsian, and M.I.
  Strikman}}(2000)}]{hrm}
\bibinfo{author}{\bibfnamefont{L.~L.} \bibnamefont{{Frankfurt, G.A. MIller,
  M.M. Sargsian, and M.I. Strikman}}}, \bibinfo{journal}{Phys.\ Rev.\ Lett.}
  \textbf{\bibinfo{volume}{84}}, \bibinfo{pages}{3045} (\bibinfo{year}{2000}).

\bibitem[{\citenamefont{{Sargsian}}(2002)}]{hrm-pre1}
\bibinfo{author}{\bibfnamefont{M.~M.} \bibnamefont{{Sargsian}}},
  \bibinfo{journal}{HEP arXiv} \textbf{\bibinfo{volume}{nucl-th/0208027}}
  (\bibinfo{year}{2002}).

\bibitem[{\citenamefont{Julia-Diaz and Lee}(2003)}]{hrm-diaz}
\bibinfo{author}{\bibfnamefont{B.}~\bibnamefont{Julia-Diaz}} \bibnamefont{and}
  \bibinfo{author}{\bibfnamefont{T.~S.~H.} \bibnamefont{Lee}},
  \bibinfo{journal}{Mod. Phys. Lett.} \textbf{\bibinfo{volume}{A18}},
  \bibinfo{pages}{200} (\bibinfo{year}{2003}).

\bibitem[{\citenamefont{{Kondratyuk \emph{et al.}}}(1993)}]{leonid}
\bibinfo{author}{\bibfnamefont{L.}~\bibnamefont{{Kondratyuk \emph{et al.}}}},
  \bibinfo{journal}{Phys.~Rev.} \textbf{\bibinfo{volume}{C48}},
  \bibinfo{pages}{2491} (\bibinfo{year}{1993}).

\bibitem[{\citenamefont{{Grishina \emph{et al.}}}(2001)}]{gris}
\bibinfo{author}{\bibfnamefont{V.~Y.} \bibnamefont{{Grishina \emph{et al.}}}},
  \bibinfo{journal}{Eur.\ Phys.\ J.} \textbf{\bibinfo{volume}{A10}},
  \bibinfo{pages}{35} (\bibinfo{year}{2001}).

\bibitem[{\citenamefont{{Grishina \emph{et al.}}}(2004)}]{gris2}
\bibinfo{author}{\bibfnamefont{V.~Y.} \bibnamefont{{Grishina \emph{et al.}}}},
  \bibinfo{journal}{Eur.\ Phys.\ J.} \textbf{\bibinfo{volume}{A19}},
  \bibinfo{pages}{117} (\bibinfo{year}{2004}).

\bibitem[{\citenamefont{Dieperink and Nagorny}(1999)}]{amec}
\bibinfo{author}{\bibfnamefont{A.~E.~L.} \bibnamefont{Dieperink}}
  \bibnamefont{and} \bibinfo{author}{\bibfnamefont{S.~I.}
  \bibnamefont{Nagorny}}, \bibinfo{journal}{Phys.\ Lett.}
  \textbf{\bibinfo{volume}{B9}}, \bibinfo{pages}{456} (\bibinfo{year}{1999}).

\bibitem[{\citenamefont{{Rossi \emph{et al.}}}(1993)}]{Rossi}
\bibinfo{author}{\bibfnamefont{P.}~\bibnamefont{{Rossi \emph{et al.}}}},
  \bibinfo{journal}{JLab\ Experiment} \textbf{\bibinfo{volume}{E93-017}}
  (\bibinfo{year}{1993}).

\bibitem[{\citenamefont{{Rossi \emph{et al.}}}(2004)}]{RossiPRL}
\bibinfo{author}{\bibfnamefont{P.}~\bibnamefont{{Rossi \emph{et al.}}}},
  \bibinfo{journal}{submitted to Phys. Rev. Lett.}  (\bibinfo{year}{2004}).

\bibitem[{\citenamefont{{Sober \emph{et al.}}}(2000)}]{tagging}
\bibinfo{author}{\bibfnamefont{D.~I.} \bibnamefont{{Sober \emph{et al.}}}},
  \bibinfo{journal}{Nucl.\ Inst.\ \& \ Meth.} \textbf{\bibinfo{volume}{A440}},
  \bibinfo{pages}{263} (\bibinfo{year}{2000}).

\bibitem[{\citenamefont{{Mecking \emph{et al.}}}(2003)}]{mecking}
\bibinfo{author}{\bibfnamefont{B.}~\bibnamefont{{Mecking \emph{et al.}}}},
  \bibinfo{journal}{Nucl.\ Inst.\ \&\ Meth.} \textbf{\bibinfo{volume}{A503/3}},
  \bibinfo{pages}{513} (\bibinfo{year}{2003}).

\bibitem[{\citenamefont{{Mirazita \emph{et al.}}}(2003)}]{Mirazita}
\bibinfo{author}{\bibfnamefont{M.}~\bibnamefont{{Mirazita \emph{et al.}}}},
  \bibinfo{journal}{CLAS Analysis Note}  (\bibinfo{year}{2003}).

\bibitem[{\citenamefont{{Crawford \emph{et al.}}}(1996)}]{Mainz}
\bibinfo{author}{\bibfnamefont{R.}~\bibnamefont{{Crawford \emph{et al.}}}},
  \bibinfo{journal}{Nucl.\ Phys.} \textbf{\bibinfo{volume}{A603}},
  \bibinfo{pages}{303} (\bibinfo{year}{1996}).

\bibitem[{\citenamefont{{Sargsian}}(2004)}]{hrm-boh}
\bibinfo{author}{\bibfnamefont{M.~M.} \bibnamefont{{Sargsian}}},
  \bibinfo{journal}{Private Communication}  (\bibinfo{year}{2004}).

\bibitem[{\citenamefont{{J. Brodsky \emph{et al.}}}(2004)}]{bfgh}
\bibinfo{author}{\bibfnamefont{S.}~\bibnamefont{{J. Brodsky \emph{et al.}}}},
  \bibinfo{journal}{Phys. Lett.} \textbf{\bibinfo{volume}{B 578}},
  \bibinfo{pages}{69} (\bibinfo{year}{2004}).

\bibitem[{\citenamefont{{H. Hiller}}(2002)}]{hiller}
\bibinfo{author}{\bibfnamefont{J.}~\bibnamefont{{H. Hiller}}},
  \bibinfo{journal}{private communication}  (\bibinfo{year}{2002}).

\end{thebibliography}

\end{document}